# Consistent Distributed Reactive Programming with Retroactive Computation

Tetsuo Kamina[a] , Tomoyuki Aotani[b] , and Hidehiko Masuhara[c]

a   Oita University, Japan
b   Sanyo-Onoda City University, Japan
c   Institute of Science Tokyo (formerly Tokyo Institute of Technology), Japan

**Abstract**

**Context**  Many systems require receiving data from multiple information sources, which act as distributed network devices that asynchronously send the latest data at their own pace to generalize various kinds of devices and connections, known as the Internet of Things (IoT). These systems often perform computations both *reactively* and *retroactively* on information received from the sources for monitoring and analytical purposes, respectively.

**Inquiry**  It is challenging to design a programming language that can describe such systems at a high level of abstraction for two reasons: (1) reactive and retroactive computations in these systems are performed alongside the execution of other application logic; and (2) information sources may be distributed, and data from these sources may arrive late or be lost entirely. Addressing these difficulties is our fundamental problem.

**Approach**  We propose a programming language that supports the following features. First, our language incorporates reactive time-varying values (also known as signals) embedded within an imperative language. Second, it supports multiple information sources that are distributed and represented as signals, meaning they can be declaratively composed to form other time-varying values. Finally, it allows computation over past values collected from information sources and recovery from inconsistency caused by packet loss. To address the aforementioned difficulties, we develop a core calculus for this proposed language.

**Knowledge**  This calculus is a hybrid of reactive/retroactive computations and imperative ones. Because of this hybrid nature, the calculus is inherently complex; however, we have simplified it as much as possible. First, its semantics are modeled as a simple, single-threaded abstraction based on typeless object calculus. Meanwhile, reactive computations that execute in parallel are modeled using a simple process calculus and are integrated with the object calculus, ensuring that the computation results are always serialized. Specifically, we show that time consistency is guaranteed in the calculus; in other words, consistency can be recovered at any checkpoint.

**Grounding**  This work is supported by formally stating and proving theorems regarding time consistency. We also conducted a microbenchmarking experiment to demonstrate that the implemented recovery process is feasible in our assumed application scenarios.

**Importance**  The ensured time consistency provides a rigorous foundation for performing analytics on computation results obtained from distributed information sources, even when these sources experience delays or packet loss.

## The Art, Science, and Engineering of Programming

## 1 Introduction

Many systems require receiving data from multiple information sources. These information sources act as distributed network devices that asynchronously send the latest data at their own pace to generalize various kinds of devices and connections, known as the Internet of Things (IoT). For example, consider a simple treadmill control system that observes the speed of the treadmill's belt and the runner's heartbeat rate. As there are various kinds of treadmills and body sensors, they are generalized as network devices that asynchronously send the latest status to the controller.

Those systems often perform computations both *reactively* and *retroactively* on information received from the sources. By reactively, we mean that the systems perform computation in response to changes of the information, which is important for monitoring purposes; for example, the treadmill system should always compare the belt speed and heartbeat rate so that it can immediately make an alert for an irregular condition. By retroactively, we mean that the systems perform computation over the series of information received in the past, which is needed for analytical purposes; for example, the treadmill system may perform statistical analysis on the record of the belt speeds and heartbeat rates for planning future exercises.

Designing a programming language that can describe such applications at a high level of abstraction, while hiding technical details from programmers, is challenging. Specifically, there are two difficulties:

- In these applications, reactive and retroactive computations are performed alongside application logic, such as starting the system or switching the display from monitoring mode to analytic mode. This application logic may include changes in the network of values that are reactively computed, which interfere with consistent retroactive computation.
- Information sources, which are updated at their own timings, may be distributed, and data from these sources may arrive late or be lost entirely.

These difficulties make existing language abstractions, such as signals in functional reactive programming (FRP) [17], unsuitable. We discuss other related work in Section 4.

To address these difficulties, this paper proposes a programming language with the following features:

**Reactive and retroactive programming with imperative operations:** Building on studies that integrate FRP-like features into imperative languages [24, 39], this language supports both reactive time-varying values (also known as signals) and imperative operations. Furthermore, leveraging techniques supported by persistent signals [23, 25, 26], this language enables computations over past values collected from information sources, which is useful for analytics. To achieve consistent retroactive computation in the presence of changes in the network of time-varying values, we introduce the mechanism of *switch history*.

**Location transparency of information sources with recovery:** This language supports multiple information sources, which may be distributed. These sources are identified by user-provided ids, making their actual locations transparent. Additionally, these sources are represented as time-varying values (in an FRP manner) and can be declaratively composed to form other time-varying values. Their update timings and synchronization methods can be specified using annotations. Furthermore, it supports recovery from inconsistencies caused by packet loss, which could result in incorrect analytics.

We note that currently no programming languages support all of these features. For example, distributed persistent signals have been proposed in the literature [26]; however, existing work does not support specifying update paces for information sources and recovery from inconsistencies.

One important research question in this paper is how time consistency (i.e., ensuring that time-series computation results do not contradict the time-series data from the information sources) is achieved in the proposed language. To answer this question, we develop the core calculus of the proposed language, $\varsigma$-DPS (distributed persistent signals), and examine the property of time consistency within the language.

Corresponding to the aforementioned language features, there are two key points in the development of this calculus:

- It requires a hybrid of reactive/retroactive computations and imperative ones. Thus, the calculus is inherently complex; however, we have simplified it as much as possible. First, the language semantics are modeled as a simple, single-threaded abstraction based on typeless object calculus [1] (tailored to persistent signals). Meanwhile, reactive computations, such as update propagations between persistent signals that execute in parallel, are modeled using a simple process calculus. Finally, these two are synchronized to form a unified calculus, ensuring that computation results are always serialized. Furthermore, to ensure that retroactive computation works properly with imperative changes in signal dependencies, the calculus models the switch history, which records dependencies between signals each time those dependencies change.

- Location transparency of information sources must be supported. To achieve this, the calculus models a directory service, which we call the ID resolver, as a mapping from identifiers of data sources to their actual database definitions. In this calculus, the recovery process is not explicitly defined but described as follows: if we rewind the computation to the time when message loss occurred and then perform the subsequent successful computations (as in the recovery process), the results are as if no message loss had occurred.

Time consistency is defined by two sub-properties: (1) the serialized computation results are always consistent with the specified update paces of the information sources and the synchronization of multiple sources, and (2) any inconsistency caused by communication failures (excluding external data sources) can always be recovered from the information sources; in other words, consistency can be restored at any checkpoint. It is guaranteed by proving two theorems, corresponding to these two sub-properties.

Our contributions are summarized as follows:

- The design and implementation of a distributed programming language that supports all the aforementioned features.

■ **Listing 1** Network monitoring example using our proposal.

```
1  @timing("every 5 sec base 00:00:00")
2  signal class Traffic {
3      persistent signal int http, https, ...;
4      signal int total = http + https + ...;
5      Traffic (String id) { ... }
6  }
7  @timing("every 1 min base 00:00:00")
8  signal class Ping {
9      persistent signal double reply;
10     signal double avg = reply.avg();
11     signal boolean dead = reply > TIMEOUT;
12     Ping (String id) { ... }
13 }
14 @timing("anytime")
15 signal class IDS {
16     persistent signal String notification;
17     IDS (String id) { ... }
18 }
19 @mode("union") @checkpointInterval(300)
20 signal class Monitor {
21     Traffic t; Ping p; IDS i;
22     signal int color = f(t.total, p.dead, i.lastTimestamp());
23     Monitor(String id, ...) { ... }
24 }
```

- The development of the simple core calculus $\varsigma$-DPS, which models both explicit application logic and implicit reactive computation.
- The foundation and proof of time consistency in $\varsigma$-DPS, ensuring that computations align with specified update pace and support recovery from communication failures.

The remainder of this paper is organized as follows. Section 2 describes the proposed language using a simple network monitoring example. Section 3 formalizes the proposed language and shows the proofs of the required properties. Section 4 discusses studies related to our proposal. Finally, Section 5 concludes the paper and discusses directions of future research.

## 2 Overview of Proposed Language

We explain the proposed language using a network monitoring example (Listing 1). This system consists of traffic monitoring (Traffic), alive monitoring (Ping), and intrusion detection (IDS) services. A traffic monitoring service monitors the amounts of packets for each protocol (e.g., http and https) and the total amount of packets (total). Similarly, an alive monitoring service and an intrusion detection service are implemented using time-series data representing the round-trip time of an ICMP packet (reply) and notifications from the IDS (notification), respectively.

Using this example, we elaborate on the problems listed in Section 1 as follows:

- This application performs both reactive and retroactive computations on data received from the sources. As a reactive computation, it triggers an alert when the monitored traffic exceeds a threshold (i.e., the color turnes red). As a retroactive computation, it provides analysis functions to understand the behavior of attackers using the past values of respective variables that are constantly updated. These reactive and retroactive computations are performed in conjunction with imperative operations, such as manually changing the service on which the traffic is monitored (not shown in Listing 1).
- The information sources providing traffic and alive monitoring, as well as intrusion detection, can be distributed across the network. These sources can be updated at their own pace. Furthermore, data from these sources may arrive late or be lost entirely, leading to incorrect analysis results based on past values.

The solution to the former problem is explained in Section 2.1, while the solution to the latter problem is provided in Section 2.2. To ensure the explanation is self-contained, the language is introduced alongside existing language mechanisms, which are presented in Sections 2.1.1 and 2.2.1. Other sections present new ideas.

## 2.1 Reactive and Retroactive Programming with Imperative Operations

This section presents the solution to the former problem among the ones listed in Section 1.

### 2.1.1 Signals and Persistent Signals

We assume that the monitoring services preexist and their monitoring results are accumulated into the corresponding time-series databases. In this language, these time-series data can be referred as signals. Signals are the principal abstraction in reactive programming (RP) languages for representing time-varying values that can be declaratively connected to form dataflows [4]. This feature is useful for implementing modern reactive systems; for example, signals can directly represent dataflows from inputs given by the environment to outputs that respond to changes in the environment. This feature was first introduced in several functional languages, such as Fran [17], and several RP extensions for imperative languages now also support this feature [24, 39]. Our language applies the latter approach; i.e., signals are *embedded* into the existing imperative language.

In our language, signals are declared using the modifier signal. For example, fields http, https, and total in class Traffic are signals (Listing 1), and total depends on http and https. This means that updates in http and https are automatically propagated to total. Thus, assuming that http and https refer to the amounts of packets of HTTP and HTTPS, respectively, which are externally updated by the data source, total always stores the total amounts of packets that are monitored by the traffic monitoring service.

The proposed language also supports *persistent signals*, which are abstractions for time-varying values with their persistent update histories [23, 25], providing an immediate way to retroactive computation. Using the modifier persistent, a persistent signal is declared as a variant of a signal that encapsulates the details of update histories.[1] For analytical purposes, these histories are queried by API methods, such as within, which returns a new persistent signal with the same history as the receiver's, but filtered using the specified time-window, and avg, which calculates the average of the receiver's persistent signal, equipped with persistent signals in advance. We may also use persistent signals for exploratory purposes such as time-travelling. Persistent signals need to be grouped into a single module called a *signal class*. For example, in Listing 1, all signals in Traffic are persistent (because signals that depend on persistent signals are also persistent) and they are grouped into the signal class (i.e., a class declared with the modifier signal) Traffic.

A signal class can be instantiated by explicitly providing the id of that instance. The formal parameter id of the constructor Traffic (as well as other constructors of signal classes) is mandatory. This instance encapsulates the dataflow (including the past values) that is identified by the id. Once created, the update history of a signal class instance is preserved on the disk, even after the application stops. Its identity is maintained on the disk, and when the application restarts, the instance identified by id becomes available again using the id as the key. The lifecycle of a signal class instance was specified in the literature [25].

### 2.1.2 Dynamic Changing of Networks

The network between signal class instances can change dynamically. One such change is an introduction of a new signal class instance performed by new; another is a replacement of upstream signal class instances, similar to the *switch* function of traditional FRP.[2] As explained below, existing mechanism of persistent signals [26] does not support this feature appropriately. In this proposal, each signal class instance implicitly provides the setUpstreams method, which replaces upstream signal class instances declared as the field of the receiver signal class instance[3] with the arguments. For example, we can switch the traffic monitoring service dynamically as follows:

```
1  m.setUpstreams(m.t, new Traffic("FileServer"));
```

We note that this dynamic changing should not be interleaved with update propagation of the connected network because, in FRP, update propagation should be atomic. To ensure this atomicity, we need to perform the following steps:

1. When changing the topology of the connected network, namely, $\mathcal{N}$, block all the source signals in $\mathcal{N}$.

---

[1] In this language, all signals annotated with persistent, as well as signals that depend on persistent signals, are persistent. Other signals are transient. In this paper, we primarily focus on persistent signals; therefore, in the subsequent sections, we consider only persistent signals.

[2] We note that each signal class instance contains a local signal network. Thus, this is a replacement of not just a single node but a sub-network.

[3] In our proposal, the upstream signal class instances must be declared as fields.

2. Then, inspect whether there are on-going propagations in $\mathcal{N}$. If so, wait until all these propagations reach the final destination in $\mathcal{N}$.

3. Finally, perform the change in $\mathcal{N}$ and unblock all sources in $\mathcal{N}$.

We note that this switch operation makes retroactive computation impossible unless the environment remembers the network at the given time, namely, t. For example, assuming that the above setUpstreams was issued at time t, the value of signal color is calculated based on the total of "FileServer" at time t1 >= t, which is calculated based on that of "WebServer" at time t2 < t. Without recovering the signal network before issuing setUpstreams, the retroactive computation at time t2 will be performed based on the signal network after issuing setUpstreams, which definitely does not result in a desired behavior.

To address this problem, we introduce the *switch history*, which is a time-series database that records the update history of the signal network. Currently, this is implemented by storing a JSON object representing a signal class instance, which is indexed with its id and the timestamp of its update, associated with the list of the id's of its upstream signal class instances. We assume that the performance impact of the introduction of this switch history should not be high in a total execution, as the switch operations are not issued frequently.

We also need to ensure that the signal networks remain acyclic; setUpstreams tests whether the resulting network is acyclic. If it would become cyclic, setUpstream warns about that and makes no changes in the signal network.

## 2.2 Location Transparency of Information Sources with Recovery

This section presents the solution to the latter problem among the ones listed in Section 1.

### 2.2.1 Distributed Persistent Signals

By introducing a directory service that maps each id to its corresponding time-series database, persistent signals can easily be realized in a distributed setting [26]. This directory service, which we call the ID resolver, is a simple key-value store, where information necessary to access the database is bound with an id (key). We assume that each application knows the location of the ID resolver, which is assumed to never fail. Each signal class instance communicates with its database and executes in the same memory space, while the database location may be distributed. It may also communicate with other signal class instances in the case of update propagation.

For example, we can instantiate Traffic, Ping, and IDS in Listing 1 using ids whose information is registered in the ID resolver. We can also instantiate Monitor, which assembles all monitoring services and updates the color indicating the status of the system (e.g., green, yellow, and red):

```
1  Monitor m = new Monitor("MyLab",
2                  new Traffic("WebServer"), new Ping("DBServer"), new IDS("FW"));
```

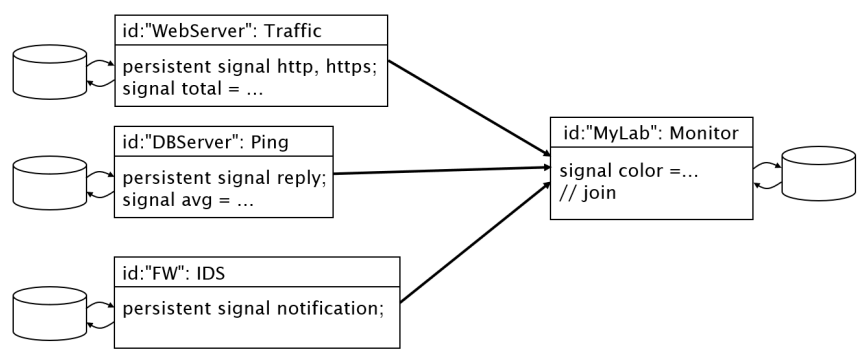

**■ Figure 1** A network of signal class instances in the network monitoring system. Each drum icon represents a database assigned to a signal class instance. These databases do not need to be local; their locations are managed by the ID resolver.

This code fragment forms a network between signal class instances. Each update (performed by the preexisting services) in databases identified by "WebServer", "DB-Server", and "FW" is propagated to the instance identified by "MyLab" (Figure 1). We note that persistent signals at the source of the dataflow are externally updatable. This update in the database is asynchronously notified to the signal class instance, and this instance then propagates it to the downstream signals in a push-manner.

This mechanism supports RP in an open and distributed setting. We can easily extend existing dataflows, whose component signal class instances are identified by public ids, by using those instances as a part of another dataflow. For example, existing dataflows can be connected to form a third-party dataflow. An existing time-series database can be considered a source of the new dataflow developed using the proposed distributed platform.

### 2.2.2  Update Timing Specification and Synchronization of Updates

Each data source may update its value at its own pace. To represent the update timing of the data source, the proposed language provides the timing annotation. This annotation expects an *update timing specification* (in the format specified below) as an argument written as a String:

    every xx (hour|min|sec) base yyyy:mm:dd:hh:mm:ss

The phrase `every xx (hour|min|sec)` means that the data source is updated every xx hours/minutes/seconds, respectively. The phase `base` specifies the time when the count of time starts. We can omit the first part yyyy:mm:dd:, which means that this update timing is reset every day at hh:mm:ss. For example, in Listing 1, every `Traffic` instance is assumed to be updated every 5 seconds from the beginning of each day. If such an update timing is unknown, we can specify it using "anytime" as an argument, as shown by the call of timing on each `IDS` instance.

If no such annotations are provided, then they are inferred from the upstream signal classes (if there are no upstream signal classes, the default annotation @timing("anytime" is applied) by calculating the update timing based on the @mode annotation explained below.

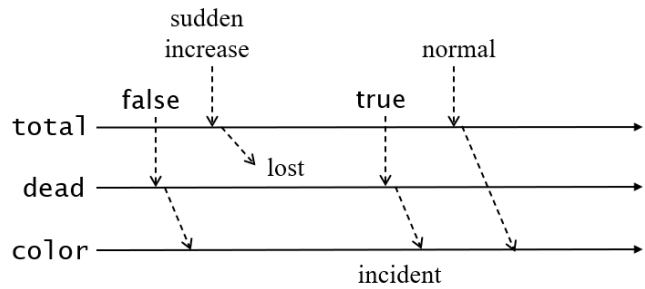

**■ Figure 2** Delayed update propagation. Dashed arrows represent propagation between persistent signals.

The `@mode` annotation (shown in the declaration of `Monitor`) indicates how to join the upstream signal class instances. If it is set to `"union"`, the recalculation occurs every time some of the depending persistent signals are updated; if it is set to `"intersection"`, the recalculation occurs only when all incoming updates are notified at the (logically) same time (the default is `"union"`). The update timing is calculated using the `@timings` of the upstream signal classes and the `@mode` of this signal class (in the case of `Monitor`, `@timing` becomes `"anytime"`).

Finally, if the signal class declares these annotations and has upstream signal classes, the compiler checks their consistency. For example, updating more frequently than the upstream signals should be rejected by the compiler.

For the correlation of incoming updates with `@timing("anytime")`, we only consider the *zip* combinator, where only the latest value of each source is used to calculate the combination. It might also be possible to consider other selection strategies of incoming updates, as proposed by Bračevac et al. [7], which is reserved for future work.

### 2.2.3 Checkpointing and Recovery

One issue that our proposal needs to consider is its time consistency. The update histories of persistent signals should be *consistent* with the data sources on which the persistent signals depend. We say that a persistent signal is consistent with the data sources if all its stored (past) values can be reproduced using the corresponding data in the data sources. Ensuring this consistency is not trivial. This is because update propagation over the network can be delayed or lost due to network latency.

For example, in Figure 2 (which refers to signals `total`, `dead`, and `color` declared in Listing 1), a sudden increase in traffic is observed on the signal `total`, followed by the observation of a `dead` host (`dead` is a signal of `boolean` representing the status of a host: `true` means that the host is possibly down; `false` means otherwise). However, the notification of updating `total` is lost, and the notification of updating `total` to be a normal amount arrives. Thus, the time-series calculations of `color` indicates that there is an incident (assuming that the dead host is considered an incident) without any anomalies in the traffic, and such incorrect time-series updates of `color` become persistent, leading to incorrect analytics of `color`, such as ignoring the fact that the cause of an incident was the sudden increase in `total`. The existing persistent signals mechanism does not provide any recovery methods for this inconsistency.

In our language, this consistency problem is tackled by performing recovery processes that are implicitly triggered by a background process at specified checkpoint timing. This checkpoint timing is specified for each signal class using the `@checkpointInterval` annotation. For example, Listing 1 shows that checkpoint processes for a `Monitor` instance are triggered every 300 seconds (i.e., 5 minutes). If this annotation is missing, no checkpoint processes are triggered for these signal class instances, and thus time consistency is not ensured.

In general, recovery to a consistent state consists of two steps: checkpointing, where each state of a transaction is recorded, and recovery, which performs the rollback [18]. However, in our setting, it is not necessary to perform rollbacks because there are no transactions; i.e., we do not have to consider aborting already performed operations, but only executing operations that have not been performed yet. Thus, our checkpointing process simply determines the checkpoint time, which is straightforward from the global clock, and represents it as a timestamp.

The recovery process in this setting is performed as follows:

1. A signal class instance annotated with `@checkpointInterval` implicitly notifies all of its most upstream signal class instances referred to by this application (i.e., signal class instances that can be updated from outside the application) about the timestamp that the checkpointing determines.

2. Each of the most upstream instances instructs all of its downstream neighbor signal class instances to start the recovery process.

3. When a signal class instance receives this instruction from all of its next upstream instances, it recalculates all its persistent signals from its "last checkpoint time" to the time specified by the timestamp determined by this checkpoint. It updates the database if necessary. After completing, it updates its "last checkpoint time" to the timestamp determined by this checkpoint, and instructs all of its downstream neighbor signal class instances to start the recovery process.

The recalculation in each signal class instance is performed by checking whether the notified signal class instance has all the records with timestamps that should exist according to the join mode specified by `@mode`:

- If the value of `@mode` is `"union"`, this signal class instance should have all records with timestamps representing the union of upstream timestamps, i.e., the union of upstream timestamps is compared with timestamps recorded on this signal class instance.

- If the value of `@mode` is `"intersection"`, this signal class instance should have all records with timestamps representing the intersection of upstream timestamps.

This recovery process ensures the consistency of the update history for each signal class instance *until the "last checkpoint time."* This is because, only upon the success of the recovery process, does the signal class instance updates the "last checkpoint time" in the database using the timestamp determined by the current checkpointing and instructs its downstream instances to start their recovery. Each instance begins its recovery after receiving these instructions from all its upstream instances. Thus, the recovery at each instance starts only after all the upstream instances successfully finish their recovery, and their histories until the current checkpointing are considered reliable, which makes the recalculation of its update history reliable.

If the recovery process at some signal class instance fails due to packet loss during the recovery or a fault in the database, its "last checkpoint time" will not be updated, and the recovery processes of all its downstream instances are blocked. The "last checkpoint time" recorded for each instance clarifies that the update history after that time is not ensured to be consistent. This possibly inconsistent status will eventually be recovered in some of the future checkpoints.

Assuming all the recovery processes are successfully performed, the time inconsistency problem discussed above is resolved. Initially, the system may report an incident without any anomalies in the traffic; however, this inconsistency is implicitly rectified by retroactively recalculating status in Figure 2 at the checkpoint timing. Consequently, the system can provide consistent analytics. The theorems in Section 3 guarantee that an inconsistent computation at any time t can be recovered by recomputations in the sense that the recomputations provide the same result as the computations as though no inconsistent computations had not occurred.

We may also consider a more strict recovery process where the consistency of the entire (connected) signal network is ensured to be recovered at a specified checkpoint time. Ideally, the recovery process should be independently defined from the language specification. We will see that the formal discussion on time consistency in Section 3.3 is independent from the applied recovery process (even though the language semantics require the blocking strategy for a more strict recovery process). As we focus on the language mechanism, we do not further detail the strict recovery process here. The microbenchmarking results shown in Appendix A show that at least the recovery process detailed above is feasible in our assumed application scenarios.

### 2.3 Premises

**Consistency model**    In this paper, we focus on consistency in retroactive computations. Persistent signals support both immediate and retroactive computations. For immediate computations, we prioritize responsiveness, an important requirement for many reactive systems, over consistency. For instance, the update propagation between signal class instances is performed using R2DBD,[4] a Reactive Streams-based variant for relational database connectivity, rather than mechanisms that strictly avoid glitches. In other words, our consistency model allows inconsistency in immediate computations, which may also be observed when a side effect such as an output from the system is triggered by immediate computation.[5] Even though this inconsistency is not observed often, the problem arises when it persists in persistent signals' update histories, which must be rectified to achieve consistency during retroactive computations.

---

[4] https://r2dbc.io/ (visited on 2024-09-16).
[5] We also consider that a non-signal computed using signals is also a side-effect.

We also assume that all source signals are updated externally; i.e., the correctness of such data sources is not handled by our proposal. In this paper, we assume that all data sources are consistently updated.

**Synchronized timestamps**  In persistent signals, every update record is assigned its timestamp. In our model, time consistency is judged based on these timestamps. This implies that we assume all timestamps in persistent signals are synchronized within a certain precision. We only consider application scenarios where the counted wall clock time, synchronized using the network time protocol [30], provides sufficient precision.

**Acyclic signal dependencies**  As in other FRP languages and systems, we assume that directed graphs constructed by signal dependencies are acyclic. In our language, this acyclicity is dynamically checked at object instantiation time.

## 3  Formalization

This section provides a formal definition of the language described in Section 2. Specifically, we discuss the guarantee of time consistency on this formalization. One problem for developing this formal definition is that, as explained in Section 2.1, there are two aspects in the proposed language: (1) the active computation on which the reactive and retroactive computation is embedded, and (2) the reactive computation that passively consumes inputs from the environment, as well as the retroactive computation that leverages the record of the reactive computation. Although there already exists a calculus for persistent signals [25], it only captures the first aspect, and the formalization of reactive update propagation is entirely missing (in fact, the value of signals in this calculus is calculated on demand). We can capture the second aspect using calculi for synchronous and FRP languages (e.g., [9]). However, we cannot capture the first aspect using such calculi alone. The first aspect is essential to discuss time consistency in this setting because some operations on persistent signals imperatively change the signal network. Thus, our key challenge is to develop a calculus that unifies these two aspects.

We designed this calculus, $\varsigma$-DPS, by representing both aspects separately using distinct semantic rules that appropriately describe its own aspect:

1. The semantics of the program are defined using a simple, single-threaded abstraction based on the object calculus [1] (tailored to accommodate persistent signals).
2. The semantics of update propagations between persistent signals, which execute in parallel, are defined using a simple process calculus, which is more suitable for representing reactions from the environment than the $\lambda$-based calculi.

To discuss time consistency, the computation results should be serialized. Thus, we then unifies them into a single calculus by synchronizing the big-step reductions of the process calculus with the small-step reductions of the object calculus. These are

$$e ::= x \mid e.p \mid e.s \mid e.m$$
$$\mid l[\overline{p} = \overline{e}, \overline{s} = \overline{e}, \overline{m} = \varsigma(\overline{x})\overline{e}] \mid$$
$$e.\text{setu}(\overline{e}) \mid l$$

$$\nu ::= \overline{l} \mapsto \overline{p}$$
$$p ::= i.\hat{l}[\overline{e}, \overline{m} = \varsigma(\overline{x})\overline{e}] \mid \hat{l}$$
$$i ::= \cup\overline{l} \mid \cap\overline{l}$$

■ **Figure 3**  Abstract syntax of $\varsigma$-DPS.  ■ **Figure 4**  Syntax of processes.

explained in Section 3.2, where we first describe how this serialization combines these two semantics, followed by the process calculus corresponding to (2) (Section 3.2.1) and the explicit reductions corresponding to (1) (Section 3.2.2). The language features explained in Section 2.2, i.e., the directory service and update timing specifications, are modeled in Section 3.1. As explained in Section 2.2.3, the recovery process is not explicitly defined in the language semantics.

We first introduce the abstract syntax of $\varsigma$-DPS based on the object calculus in Figure 3. Let the metavariable p range over persistent signals; e range over expressions; x range over variables, which represents *self* in the object calculus; l range over identifiers; and m range over method names. For simplicity, we adopt the convention invented by FJ [21] to represent sequences, i.e., overlines denote sequences. For example, $\overline{e}$ represents a possibly empty sequence $e_i, \cdots, e_n$, where $n$ denotes the length of the sequence. We use $\overline{p}=\overline{e}$ as shorthand for "$p_1=e_1 \cdots p_n=e_n$," and $\overline{m} = \varsigma(\overline{x})\overline{e}$ as shorthand for "$m_1 = \varsigma(x_1)e_1, \cdots, m_n = \varsigma(x_n)e_n$."

An expression can be either a variable, access to either a persistent signal, upstream signal class instance, or method, object labeled with an identifier, invocation of setu (an abbreviation of setUpstreams), or an identifier l. We assume the set *Id* of identifiers and $l \in Id$. We note that the syntax of object is slightly modified from the original object calculus. First, we distinguish methods from other fields such as persistent signals and upstream signal class instances in an object. Second, an object is labeled with the identifier l, which also acts as a reference to the object. This means that $\varsigma$-DPS supports references. Unlike the object calculus, $\varsigma$-DPS does not support overriding. This is because subclassing actually does not interact with the behavior of signal class instances, which are characterized by persistent signals, making them orthogonal to the method lookup.

This ignorance of overriding makes the type system uninteresting; thus, $\varsigma$-DPS is designed as an untyped calculus. Furthermore, it is not class-based, even though the proposed language is class-based. Assuming the following signal class declaration

```
1  mode("union") timing(30) signal class C{
2     signal p=l.p₁; s; C(id,s){ this.s=s; } m(){ e; }
3  }
```

the signal class instance new $C(l_0, l_1)$ is encoded into the object $l_0[p = l.p_1, s = l_1, m = \varsigma(x)e]$ (recall that x is not a formal parameter of m, but represents *self*). As discussed below, we assume that the database relation corresponding to $l_0$ is provided a priori. Similarly, the information provided by @mode and @timing is provided by an environment representing the ID resolver.

### 3.1 ID Resolver and Databases

We first model the ID resolver as the proposed language relies on it. The ID resolver is formalized as the identifier environment $\mu$, which is a set of mapping $l \mapsto (\mathscr{R}_l, tm, mode)$, where $l$ is the identifier of the signal class instance and $\mathscr{R}_l$ is an execution history of the object identified by $l$. The schema of this execution history is defined as $schema(\mathscr{R}_l) = (time, \overline{p})$. The attribute name $time$ represents timestamps $t \in Time$, where $Time$ is a total order set where $\bot \in Time$ and $\forall t \in Time.\ \bot \le t$, i.e., $\bot$ is a bottom element. For convenience, in the following, we assume that $Time$ is a set of non-negative integers. The attribute names $\overline{p}$ represent persistent signals contained in the object $l[\overline{p} = \overline{e}, \cdots]$.

Intuitively, $\mu$ models the ID resolver, which is a simple key-value store that maps an identifier $l$ to its corresponding database $\mathscr{R}_l$, update timing $tm$, and the synchronization mode $mode$. For simplicity, an update timing $tm$ is an integer corresponding to "every $tm$ base $\bot$." The $mode$ can be either $\cup$ (the join is calculated every time some of the depending persistent signals are updated) or $\cap$ (the join is calculated only when all incoming updates are notified at the same time). We assume that the ID resolver is prepared in advance. Therefore, in this model, we assume that the database relation mapped from $l$ is provided a priori. For example, $\mu$ has the following entry for the above object identified by $l_0$:

$$l_0 \mapsto (\mathscr{R}_{l_0}, 30, \cup)$$

where the schema of $\mathscr{R}_{l_0}$ is defined as $(time, p)$.

Each relation is handled using the operations provided by the relational algebra [11]: $\pi_{col}(\mathscr{R})$ represents a projection of a relation $\mathscr{R}$ by $col$ (i.e., selecting the column $col$ from $\mathscr{R}$), and $\sigma_c(\mathscr{R})$ represents filtering $\mathscr{R}$ using the condition $c$. We often use a singleton set $\{l\}$ and its value $l$ interchangeably. As a condition $c$, we can use a predicate $latest$, which is true only if the $time$ field of the tuple has the largest value among the relation, or $latest(t)$, which is defined as $\sigma_{latest}(\sigma_{\le t})$. We assume that every relation $\mathscr{R}$ contains a tuple whose value of $time$ is $\bot$; i.e., $\sigma_{latest(\bot)}(\mathscr{R})$ is always defined.

### 3.2 Runtime Semantics

To discuss time consistency, we need to express time when the computation is taking place. To specify time, we use timestamps, and as we already introduced, we use the metavariable $t$ that ranges over timestamps.

Besides $\mu$ and $t$, there are additional environments that are required to express our formal semantics. The process environment $\nu$ represents the current state of the program and is defined as a mapping from identifiers to objects (we will see why it is called "process environment" shortly). The switch history $\phi$ is a set of pairs $(t, \nu)$, where $\nu$ is a program state at time $t$. It formalizes the switch record discussed in Section 2.1.2. Then, the reduction rules for expressions are given in the form $\mu; t \vdash \nu; \phi \mid e \longrightarrow \nu'; \phi' \mid e'$, which is read as "under an identifier environment $\mu$ at time $t$, an expression $e$ with process environment $\nu$ and switch history $\phi$ reduces to $e'$ with $\nu'$ and $\phi'$."

$$\frac{\mu;\nu;\mathsf{t}\vdash_{p(\overline{l'})}\overline{e}\longrightarrow^*\overline{l'}\qquad l\mapsto(\mathscr{R}_l,tm,mode)\in\mu}{\mathsf{t}\ mod\ tm=0\qquad\mathscr{R}'_l=\mathscr{R}_l\oplus\{(\mathsf{t},\overline{l'})\}\qquad\mu'=\mu\oplus l\mapsto(\mathscr{R}'_l,tm,mode)}{\nu;\mathsf{t}\vdash\mu\mid_{\overline{l'}}\cdots,l_i\mapsto\hat{l_i},l\mapsto\cup\overline{l}.\hat{l}[\overline{e},\overline{m}=\varsigma(\overline{x})\overline{e'}]\leadsto\mu'\mid_{l,\overline{l'}}\cdots,l_i\mapsto\hat{l_i},l\mapsto\hat{l}}$$
$$(\textsc{Pr-Or})$$

$$\frac{\mu;\nu;\mathsf{t}\vdash_{p(\overline{l'})}\overline{e}\longrightarrow^*\overline{l'}\qquad l\mapsto(\mathscr{R}_l,tm,mode)\in\mu}{\mathsf{t}\ mod\ tm=0\qquad\mathscr{R}'_l=\mathscr{R}_l\oplus\{(\mathsf{t},\overline{l'})\}\qquad\mu'=\mu\oplus l\mapsto(\mathscr{R}'_l,tm,mode)}{\nu;\mathsf{t}\vdash\mu\mid_{\overline{l'}}\cdots,\overline{l}\mapsto\hat{\overline{l}},l\mapsto\cap\overline{l}.\hat{l}[\overline{e},\overline{m}=\varsigma(\overline{x})\overline{e'}]\leadsto\mu'\mid_{l,\overline{l'}}\cdots,\overline{l}\mapsto\hat{\overline{l}},l\mapsto\hat{l}}$$
$$(\textsc{Pr-And})$$

■ **Figure 5** Reduction of processes.

Our basic ideas to unify the update propagations and other computations (which we call explicit reductions here) are two fold: (1) time proceeds after one of the explicit reductions takes place, and (2) if update propagations occur at time t, they are taken place between the explicit reductions. Assuming the propagation is given by the rule in the form $\nu;\mathsf{t}\vdash\mu\longrightarrow\mu'$ (read "under a process environment $\nu$ at time t, $\mu$ is updated to $\mu'$"), this strategy is represented by the following single time-step, written $\mu;\nu;\phi;\mathsf{t}\mid e\twoheadrightarrow_{\overline{\mathsf{t}}}\mu';\nu';\phi';\mathsf{t}+1\mid e'$, which is defined as a concatenation of a propagation and an explicit reduction:

$$\frac{\mu;\mathsf{t}\vdash\nu;\phi\mid e\longrightarrow\nu';\phi'\mid e'\qquad\nu;\mathsf{t}\vdash\mu\longrightarrow_{\overline{\mathsf{t}}}\mu'}{\mu;\nu;\phi;\mathsf{t}\mid e\twoheadrightarrow_{\overline{\mathsf{t}}}\mu';\nu';\phi';\mathsf{t}+1\mid e'}\qquad(\textsc{R-Step})$$

Here, t represents logical time, and $\nu;\mathsf{t}\vdash\mu\longrightarrow\mu'$ represents update propagations occurred between explicit reductions; these propagations produce effects only on $\mu$. Time is updated if an explicit reduction proceeds under the environment $\mu'$ updated by the propagations. To precisely define the propagation rule, we provide the semantics of update propagation in the form of process calculus, which is detailed as follows.

### 3.2.1 Processes

To represent update propagations between persistent signals, we consider each object as a process, which receives messages from objects corresponding to the upstream signal class instances. To make this explicit, we slightly modify the syntax of objects, which we call processes here. The syntax of processes is presented in Figure 4. A process $p$ has a join $i$ of input channels $\overline{l}$, which guards emitting propagation through an output channel $\hat{l}$, or consists of only an output channel. Intuitively, these input channels listen to the messages from the upstream signal class instances. Thus, they are identified by the identifiers for them. The output channel identifies the process itself. The list of expressions $[\overline{e},\cdots]$ is taken from the right-hand side of signals $\overline{p}=\overline{e}$ in the object, and represents effects of this process; when emitting the propagation, these expressions reduce to values $\overline{l}$ and are inserted into the corresponding database in $\mu$.

A join $i$ can be either $\cup\bar{l}$ (this guard is removed if one of the input channels is emitted) or $\cap\bar{l}$ (this guard is removed if every input channel is emitted). For example, the process of the object identified by $l_0$ shown above is written as $\cup l_1\hat{l_0}[l.p_1, m = \varsigma(x)e]$. The process environment $\nu$ is a list of mappings $l \mapsto p$ from an identifier $l$ to the process $p$ identified by $l$. In the process reduction explained below, $\nu$ is viewed as a parallel composition of the processes. As mentioned above, $\nu$ is also viewed as a state of the program on which an object identified by $l$ is searched (see R-Invk below). This is why a process includes the list of methods $\bar{m}$.

The behavior of the processes is given by the process reduction of the form $\nu; t \vdash \mu \mid_{\bar{l}} \nu' \rightsquigarrow \mu' \mid_{\bar{l'}} \nu''$. In the process reduction, we view $\nu$ as a parallel composition of the processes in the range of $\nu$; $\nu'$ on the left-hand side of $\rightsquigarrow$ is the parallel composition before the update propagation, and $\nu''$ on the right-hand side is the parallel composition after the update propagation. Similarly, $\mu$ in the left-hand side of $\rightsquigarrow$ is the environment before the update propagation, and $\mu'$ in the left-hand side is the environment that is modified by the propagation. $\bar{l}$ and $\bar{l'}$ are list of identifiers of processes that have been reduced during the propagations (these are provided in the rules simply for the convenience of proof construction). The left-hand side of $\vdash$ indicates the process environment and time when the propagation is initiated.

The main rules for process reductions are shown in Figure 5 (as in most process calculi, the parallel composition is commutative). Each rule consumes output channels to remove the guard of the process. The rule Pr-Or is applied when the guard is $\cup\bar{l}$, which consumes the output channel $\hat{l_i}$, where $l_i \in \bar{l}$, placed in the parallel composition. The rule Pr-And is applied when the guard is $\cap\bar{l}$, which consumes all the output channels $\hat{\bar{l}}$ in the parallel composition. Other conditions and effects are identical to each rule:

- Each rule is applied only when $t \ mod \ tm = 0$. Unlike Section 2.2.2, $tm$ here represents update pace in logical time. Without loss of generality, we can abstract the difference between the wall clock and logical time here.

- The data source $\mathcal{R}_l$ and update timing $tm$ corresponding to $l$ are retrieved from $\mu$ (as indicated by $l \mapsto (\mathcal{R}_l, tm, mode) \in \mu$)

- As an effect, expressions $\bar{e}$ (attached to $l$) are evaluated to values $\bar{l'}$ using $\mu$ (the identifier environment before the rule is applied) and $\nu$ (the process environment at the time the propagation is initiated). This effect is indicated by $\mu; \nu; t \vdash_p \bar{e} \longrightarrow^* \bar{l'}$ in the premises of the rules; the pure reduction of the form $\mu; \nu; t \vdash_B e \longrightarrow e'$ will be introduced shortly.[6]

- $\bar{l'}$ are inserted into the relation $\mathcal{R}_l$ with the timestamp $t$, resulting in a new environment $\mu''$ (as indicated by $\mathcal{R}_l' = \mathcal{R}_l \oplus \{(t, \bar{l'})\}$ and $\mu'' = \mu' \oplus l \mapsto (\mathcal{R}_l', tm, mode)$).

---

[6] The pure reduction rules $\mu; \nu; t \vdash e \longrightarrow e'$ are not affected by the differences between $\nu$ under the ongoing propagation. In this sense, $\nu$ in the left-hand side of the process reduction rules is not necessary; we can use the process environment at the left-hand side of $\rightsquigarrow$ in the left-hand side of $\vdash$ of the pure reduction. We simply adopt this redundancy to avoid making the process reduction rules too verbose.

$$\frac{\mu;\nu;t \vdash_b e \longrightarrow e'}{\mu;t \vdash \nu;\phi \mid e \longrightarrow \nu;\phi \mid e'} \qquad \text{(R-Pure)}$$

$$\frac{\mu(l) = (\cdots, mode) \qquad \nu' = \nu \oplus l \mapsto mode \; \bar{l}.\hat{l}[\overline{e},\overline{m} \mapsto \varsigma(\overline{x})\overline{e}]}{\mu;t \vdash \nu;\phi \mid l[\overline{q} = \overline{e}, \overline{s} = \overline{l}, \overline{m} = \varsigma(\overline{x})\overline{e}] \longrightarrow \nu';\phi \cup \{(t,\nu')\} \mid l} \qquad \text{(R-Obj)}$$

$$\frac{\nu(l) = mode \; \bar{l'}.\hat{l}[\cdots] \qquad \nu' = \nu \oplus l \mapsto mode \; \bar{l}.\hat{l}[\cdots]}{\mu;t \vdash \nu;\phi \mid l.\text{setu}(\bar{l}) \longrightarrow \nu';\phi \cup \{(t,\nu')\} \mid l} \qquad \text{(R-Setu)}$$

■ **Figure 6** Reduction rules for expressions.

In short, each process reduction rule represents an internal behavior of a signal class instance that updates the relation $\mathscr{R}_l$ using the final results of $\overline{e}$.

We then define the propagation (written as $\nu;t \vdash \mu \longrightarrow \mu'$) throughout an entire network of signal class instances $\nu$ in a big-step manner:

$$\frac{\nu;t \vdash \mu \mid_\emptyset \; \nu \leadsto^* \mu' \mid_{\bar{l}} \nu' \qquad \bar{l} = \{l \in dom(\mu) \mid t \equiv 0 \; mod \; tm(\mu(l))\}}{\nu;t \vdash \mu \longrightarrow_{\bar{l}} \mu'} \qquad \text{(Propagation)}$$

This means that propagation is an atomic computation where $\nu$ is reduced to $\nu'$ using multiple process reductions, and all $\mathscr{R}_l$ in $\mu$ that should be updated are updated (as indicated by the second premise of Propagation).

The propagation always starts by consuming the process with the empty guard. This is because, according to Pr-Or and Pr-And, a non-empty guard always requires emission of output channels to reduce, and there are no emitted output channels in $\nu$ in the Propagation rule. Furthermore, it is apparent that a process is never evaluated twice in one entire propagation, because there are no rules to reduce unguarded processes. As the propagation only updates the databases, the network of signal class instances $\nu$ does not change. An example of process reduction is shown in Appendix B. Even though this Propagation rule updates all processes at once, this does not contradict the situation where updates are not performed atomically over the whole network, unless the latecomer updates overtake the preceding ones.

### 3.2.2 Explicit Reduction

We show the reduction rules, given by the relation of the form $\mu;t \vdash \nu;\phi \mid e \longrightarrow \nu';\phi' \mid e'$, of expressions in Figure 6. We note that the reduction rules include R-Pure, meaning pure reductions, which use the rules defined in Figure 7. This separation of rules is necessary because the pure reduction rules are also referred to by the process reductions shown in Figure 5. In other words, non-pure reduction rules, which modify either $\mu$ or $\nu$, cannot be applied during the update propagation.

The rule R-Obj defines the reduction of the object; an object reduces to its identifier $l$. It also adds the mapping from $l$ to its process to $\nu$. We use $\oplus$ as a destructive update

$$\frac{\mu(l_0) = (\mathscr{R}_{l_0}, \cdots) \qquad \pi_p(\sigma_{latest(t-1)}(\mathscr{R}_{l_0})) = \{l\}}{\mu; \nu; t \vdash_b l_0.p \longrightarrow l}$$

(R-SignalBuild)

$$\frac{\mu(l_0) = (\mathscr{R}_{l_0}, \cdots) \qquad \pi_p(\sigma_{latest(t)}(\mathscr{R}_{l_0})) = \{l\} \qquad l_0 \in \bar{l}}{\mu; \nu; t \vdash_{p(\bar{l})} l_0.p \longrightarrow l}$$

(R-SignalPropagate1)

$$\frac{\mu(l_0) = (\mathscr{R}_{l_0}, \cdots) \qquad \pi_p(\sigma_{latest(t-1)}(\mathscr{R}_{l_0})) = \{l\} \qquad l_0 \notin \bar{l}}{\mu; \nu; t \vdash_{p(\bar{l})} l_0.p \longrightarrow l}$$

(R-SignalPropagate2)

$$\frac{\nu(l_0) = mode \; \bar{l}.\hat{l_0}[\cdots]}{\mu; \nu; t \vdash_B l_0.s_i \longrightarrow l_i}$$

(R-SignlCls)

$$\frac{\nu(l_0) = mode \; \bar{l}.\hat{l_0}[\cdots, m = \varsigma(x)e, \cdots]}{\mu; \nu; t \vdash_B l_0.m \longrightarrow e[x/l_0]}$$

(R-Invk)

■ **Figure 7** Pure reductions.

of the mapping (note that the left-hand side of $\oplus$ expects a single element, while the right-hand side expects a set), i.e., $x \oplus y = x$ if $dom(y) \in dom(x)$; $x \oplus y = x \cup \{y\}$ otherwise. This process is constructed by referring to its synchronization mode stored in $\mu$ to determine the form of the input channels. As this rule modifies $\nu$, the switch history $\phi$ is also updated.

The rule R-Setu for the call of **setu**, which represents the replacement of upstream signal class instances, updates $\nu$ using the given arguments $\bar{l}$ accordingly. We assume that this update maintains the signal network as acyclic. It also updates the switch history $\phi$.

**Pure reduction** Pure reductions are used in both propagation and explicit reductions, requiring careful handling when reading values from a persistent signal. In R-Step, explicit reduction takes place before propagation. As a result, the latest values of persistent signals referenced during explicit reduction are those from before propagation. For instance, if propagation occurs at time t, and R-Step is retroactively executed at t, the explicit reduction must not access persistent signal values at t. To address this issue, the pure reduction rules are annotated with their evaluation mode, specifying whether the pure reduction is applied in the context of explicit reduction or propagation.

Pure reductions are given by the relation of the form $\mu; \nu; t \vdash_B e \longrightarrow e'$, where $B$ is a meta variable representing the evaluation mode: $p(\bar{l})$ (propagation, where $\bar{l}$ are already computed latest values during this propagation) or $b$ (explicit reduction) (Figure 7). There are four rules for access to the members of objects according to the types of labels. The rule R-SignalBuild defines how an access to a persistent signal p behaves, which results in the value in column p of the latest tuple in $\mathscr{R}_l$ from before propagation at t (indicated by $latest(\mathsf{t}-1)$). The rule R-SignalPropagate also defines an access to a persistent signal, but from propagation. This rule is further split for two subcases: if the receiver process $\mathsf{l}_0$ has already been computed during the propagation, the value of persistent signal results in the one computed during the propagation at t; otherwise, the value results in the one before the propagation. The rule R-SgnlCls defines reduction of an access to its upstream signal class instance, which results in the identifier of that instance as found in $v$. R-Invk defines reduction of an access to a method, which results in the body of the method, where the parameter x (representing *self* in the object calculus) is replaced with the receiver of the method access.

**Congruence** We also define the congruence rule that enables a reduction of subexpressions. For this purpose, we first introduce the evaluation context $E$, which is defined as follows:

$$E ::= [\,] \mid \mathsf{l}[\overline{\mathsf{p}} = \overline{\mathsf{e}}, \overline{\mathsf{s}} = \overline{\mathsf{l}}, \mathsf{s} = E, \overline{\mathsf{s}} = \overline{\mathsf{e}}, \overline{\mathsf{m}} = \varsigma(\overline{\mathsf{x}})\overline{\mathsf{e}}] \mid E.\mathsf{setu}(\overline{\mathsf{e}}) \mid \mathsf{l}.\mathsf{setu}(\overline{\mathsf{l}}, E, \overline{\mathsf{e}})$$

Each evaluation context is an expression with a hole (written as $[\,]$) somewhere inside it. We write $E[\mathsf{e}]$ for an expression obtained by replacing the hole in $E$ with $\mathsf{e}$.

Using $E$, the congruence rule is defined as follows:

$$\frac{\mu; \mathsf{t} \vdash v; \phi \mid \mathsf{e} \longrightarrow v'; \phi' \mid \mathsf{e}'}{\mu; \mathsf{t} \vdash v; \phi \mid E[\mathsf{e}] \longrightarrow v'; \phi' \mid E[\mathsf{e}']} \tag{R-Cngr}$$

The evaluation context syntactically defines the evaluation order of subexpressions in a method invocation, e.g., the arguments are not reduced until the receiver becomes an identifier.

Similarly, we define the congruence rule for pure reduction using the evaluation context:

$$E ::= [\,] \mid E.\mathsf{p} \mid E.\mathsf{s} \mid E.\mathsf{m}$$

The congruence rule for pure reductions is defined as follows:

$$\frac{\mu; v; \mathsf{t} \vdash_B \mathsf{e} \longrightarrow \mathsf{e}'}{\mu; v; \mathsf{t} \vdash_B E[\mathsf{e}] \longrightarrow E[\mathsf{e}']} \tag{R-CngrPure}$$

An example of the whole reductions is shown in Appendix B.

### 3.3 Time Consistency

We discuss the time consistency of our proposal, that is, we can go back in time in the update propagation history of the dataflow consistently. This property is defined based on the consistency of the computation results (stored in $\mu$) with respect to the user-provided specifications, which are provided in the form of $tm$ and $mode$ in $\mu(l)$. Another important aspect of time consistency is that the lost communications (due to packet losses) can always be recovered from the sources of the dataflow. Below, we present the theorems that formally represent these properties. Proofs are presented in Appendix C.

To discuss the consistency with respect to the user-provided specifications, we first need to check whether $\mu$ is consistent with respect to the user-provided specifications. For each $l_0 \in dom(\mu)$, we can check this condition by the following judgment ($gcd$ and $lcm$ represent greatest common divisor and least common multiplier, respectively):

$$\frac{\mu(l_0) = (\cdots, tm, mode) \qquad \nu(l_0) = mode\,\bar{l}.l_0^{\hat{}}[\cdots] \qquad \mu(\bar{l}) = (\cdots, \overline{tm}, \cdots)}{(mode = \cup \wedge tm = gcd(\overline{tm})) \vee (mode = \cap \wedge tm = lcm(\overline{tm}))}$$
$$\mu, \nu \vdash l_0 \text{ OK}$$

Intuitively, this judgment checks whether the user-provided specifications on $\bar{l}$, which are $l_0$'s immediate upstream processes, do not contradict the user-provided specification on $l_0$. If $l_0$'s mode is $\cup$, its $tm$ should be the greatest common divisor of $\bar{l}$'s $tms$. Otherwise, the $tm$ should be the least common multiple of $\bar{l}$'s $tms$. A process environment $\nu$ is said to be wellformed under $\mu$ if $\forall l \in dom(\nu).\mu, \nu \vdash l$ OK and $dom(\nu) \subseteq dom(\mu)$.

A database relation is said to be consistent if all tuples in the relation are inserted as expected by the user-provided synchronization mode, which is defined as follows:

**Definition 3.1** (Consistent Relation). *Let $\iota(p)$ be input channels of a process $p$, $\mathscr{R}(\mu(l))$ be the relation of $\mu(l)$, and $mode(\mu(l))$ be the mode of $\mu(l)$. An identifier environment $\mu$ is consistent under $\nu$ at $t$ and $l$ if*

1. *If $mode(\mu(l)) = \cap$, then $\exists \bar{l}.(t, \bar{l}) \in \mathscr{R}(\mu(l)) \iff \forall l_i \in \iota(\nu(t)(l)).\exists \bar{l'}.(t, \bar{l'}) \in \mathscr{R}(\mu(l_i))$.*
2. *If $mode(\mu(l)) = \cup$, then $\exists \bar{l}.(t, \bar{l}) \in \mathscr{R}(\mu(l)) \iff \exists l_i \in \iota(\nu(t)(l)).\exists \bar{l'}.(t, \bar{l'}) \in \mathscr{R}(\mu(l_i))$.*

An identifier environment $\mu$ is said to be consistent under $\nu$ at $t$ if $\forall l \in dom(\mu).\mu$ is consistent under $\nu$ at $t$ and $l$. The following theorem guarantees that the update propagation throughout $\nu$ that is wellformed under $\mu$ preserves the user-provided specifications if setu does not break the wellformedness of the process environment[7] ($\phi(t)$ is a simplification of $\phi(max\{t' \in dom(\phi)|t' <= t\})$).

**Theorem 3.1** (Consistency of Update Timing). *If $\nu$ is wellformed under $\mu$, $\phi(t) = \nu$, $\forall l \in dom(\mu).(\exists \bar{l'}.(t, \bar{l'}) \in \mathscr{R}(\mu(l))) \implies t \equiv 0 \bmod tm(\mu(l))$, and $\mu; \nu; \phi; t \mid e \twoheadrightarrow \mu'; \nu'; \phi'; t+1 \mid e'$, then $\mu'$ is consistent under $\phi'(t)$ at $t$.*

To discuss the recovery from the case of lost communications, we first need to clarify what lost communications are, because the calculus shown above does not model them. Our calculus can easily be extended to model a lost communication, which is an update propagation that does *not* result in a normal form. We define a lost propagation as follows:

$$\frac{\nu;t \vdash \mu \mid_\emptyset \; \nu \rightsquigarrow^* \mu' \mid_{\bar{l}} \nu' \qquad \bar{l} \subseteq \{l \in dom(\mu) \mid t \equiv 0 \; mod \; tm(\mu(l))\}}{\nu;t \vdash \mu \longrightarrow_{\bar{l}} \mu'}$$

---

[7] This assumption should be satisfied by the proposed language because the update timing is specified for the class; i.e., the replacing signal class instance and the replaced one share the same update timing.

(PropLost)

The lost propagation variant of R-Step, written $\mu;\nu;\phi;t \mid e \twoheadrightarrow_{-\bar{l}} \mu';\nu';\phi';t+1 \mid e'$, is also defined accordingly:

$$\frac{\mu;t \vdash \nu;\phi \mid e \longrightarrow \nu';\phi' \mid e' \qquad \nu;t \vdash \mu \longrightarrow_{\bar{l}(\mathrm{PropLost})} \mu'}{\mu;\nu;\phi;t \mid e \twoheadrightarrow_{-\bar{l}} \mu';\nu';\phi';t+1 \mid e'}$$

(R-StepLost)

The recovery from the lost propagation is performed at some specified checkpoint. As discussed in Section 2.2.3, there can be multiple recovery methods, and the language semantics can be independent from the applied recovery method. Thus, in ς-DPS, neither the checkpoint and recovery process are not modeled explicitly. Because the recovery is a process that recomputes the persistent signals at some specific time, only the property we require is that, this actually recovers the persistent signals as if there was no lost propagation. This means that, if we rewind the computation to the time when R-StepLost was taken (namely, t) and then perform R-Step and the subsequent reductions, the results are compatible with the computation above, except that at time t R-Step was taken instead of R-StepLost. This property can be guaranteed by proving the following theorem. Suppose that $e$ under $\mu$, $\nu$, and $\phi$ at time t is evaluated to $e'$ with one step, changing environments and the switch history to $\mu^1$, $\nu^1$, and $\phi^1$ at time $t^1$. If this step is performed under lost propagation, the results might be different from $e'$, $\mu^1$, $\nu^1$, and $\phi^1$. Let these different results be $e''$, $\mu^2$, $\nu^2$, and $\phi^2$. The recovery process re-executes $e$ using these identifier environment (i.e., $\mu^2$) and switch history (i.e., $\phi^2$), however using the process environment at time $t-1$ (i.e., $\phi^2(t-1)$) (viewing $\phi$ as a relation whose schema is $(time, \nu)$, $\phi(t) = \pi_\nu(\sigma_{latest(t)}(\phi))$). Basically, the recovery theorem states that this re-execution results in $e'$ under $\mu^1$, $\nu^1$, and $\phi^1$ at time $t^1$, which are the same as the supposed results.

To formally state this theorem, we use the following notations:

- $\mu \overset{\mathscr{R}^{\leq t}}{=} \mu'$ if $\forall l \in dom(\mu).\forall t'.t' \leq t \implies \exists \bar{l}.((t,\bar{l}) \in \mathscr{R}(\mu(l)) \iff (t,\bar{l}) \in \mathscr{R}(\mu'(l)))$
- $\mu \overset{tm}{=} \mu'$ if $\forall l \in dom(\mu).tm(\mu(l)) = tm(\mu(l'))$
- $\mu \overset{mode}{=} \mu'$ if $\forall l \in dom(\mu).mode(\mu(l)) = mode(\mu(l'))$

The recovery theorem is written as:

**Theorem 3.2** (Recovery). *Suppose that* $\forall t' \in dom(\phi).t' < t$, $\phi(t) = \nu$, $\forall l \in dom(\mu).\forall t' \geq t$. $/\exists \overline{l''}.(t',\overline{l''}) \in \mathscr{R}(\mu(l))$, $\mu;\nu;\phi;t \mid e \twoheadrightarrow_{\bar{l}} \nu^1;\phi^1;t^1 \mid e'$, *and* $\mu;\nu;\phi;t \mid e \twoheadrightarrow_{-\overline{l'}} \mu^2;\nu^2;\phi^2;t^2 \mid e''$. *Then, there exists* $\mu^3$, $\nu^3$, $\phi^3$, $t^3$, *and* $e'''$ *such that* $\mu^2;\phi^2(t-1);\phi^2;t \mid e \twoheadrightarrow_{\bar{l}} \mu^3;\nu^3;\phi^3;t^3 \mid e'''$, $\mu^1 \overset{\mathscr{R}^{\leq t}}{=} \mu^3$, $\mu^1 \overset{tm}{=} \mu^3$, $\mu^1 \overset{mode}{=} \mu^3$, $\nu^1 = \nu^3$, $\phi^1 \overset{\leq t}{=} \phi^3$, $e' = e'''$, *and* $t^1 = t^3$.

## 4  Related Work

**FRP and clocks**   Reactive language features such as update propagations has been studied in several synchronous languages such as LUSTRE [19], ESTEREL [6], and LUCID SYNCHRONE [36]. These language designs are inspired by Kahn processes [22], and formal semantics are given by synchronous Kahn networks [9] and their extensions [12], which guarantees timeleak freedom by a type system considering a clock as a type. FRP languages provides a richer model than these languages do [3, 27]. These formalizations mostly focus on reactive language features, which include *switch*. On the other hand, signals are also studied in object-oriented languages, where signals are *harmonized* with imperative language features [24, 25]. For example, an effect similar to FRP's switch is produced by an imperative field set (the setUpstream in this paper also behaves in this manner). Core calculi for these proposals were developed based on FJ [21], and in these calculi, values of signals are simply evaluated in a pull-based manner. However, push-based calculation is inevitable in a distributed environment, making the discussion regarding consistency between distributed time-series data is necessary. ς-DPS differs from these pieces of work in that it can handle this discussion in a setting of distributed object-oriented language by making it as a hybrid of process calculus and object calculus.

Rhine [5] is a DSL embedded in Haskell that coordinates reactive processes at different update rates. In Rhine, data and clocking aspects, as well as synchronous and asynchronous aspects, are separated and can be specified, and clock information is expressed at the type level so that ill-clocked programs are rejected at compile time. EvEmfrp [43], which is an extension of Emfrp [40], an FRP language designed for small-scale embedded systems, also provides the language constructs to describe periodic and aperiodic tasks that are coordinated with each other. In general, these languages work in a situation where data are updated thousands of times per second. Thus, our proposed recovery method at each checkpoint time may not work in a worst case where recovery takes tens of milliseconds. Instead, our language coordinates reactive processes in a distributed setting, and to our knowledge, these existing pieces of work do not provide the feature of retroactive computation.

**Distributed and fault-tolerant RP**   Our proposal is related to constructing distributed systems using RP languages; i.e., it is closely related to distributed RP. Glitches make the realization of distributed RP challenging. Ensuring glitch freedom is not difficult in a local system, because we can determine the order of updates according to the form of dataflows by applying topological sorting [29]. However, this is challenging to perform in a distributed setting. First, we need to consider network latency and faults in distributed nodes. Second, it is not trivial who determines the order of updates through the distributed dataflows. To address this issue, several proposals have been presented; e.g., attaching versions to time-varying values [28, 42], introducing a coordinator with a responsibility to determine the update order [16], a more generalized form or variants of distributed glitch-freedom [14], and a distributed algorithm that determines the update order by exchanging information among the distributed nodes [33]. All these proposals consider signals that are transient in a distributed setting.

Conversely, this study aimed to make persistent signals distributed and consistent. Because a persistent signal records its update history with timestamps, we proposed using these timestamps to ensure consistency throughout the distributed dataflows. In this study, we assume these timestamps are synchronized. To realize this assumption, we can rely on the existing time synchronization mechanisms such as the network time protocol, which we consider provides enough precision in many cases. The time consistency of our proposal implies its glitch-freedom *in past computations*. Our system prioritize responsiveness in the immediate computation, allowing temporal inconsistency; however, this inconsistency is recovered at the specified checkpoint time.

We note that QPROP [33] also supports eventual consistency by propagating only updates of fully matched incoming signals that are time consistent. In QPROP, inconsistencies are determined solely by the incoming messages, making recovery processes triggered at regular intervals unnecessary. By adapting QPROP to consider different modes, such as @union and @intersection, we could take a similar approach, leveraging the timestamps of messages to identify dropped messages. This direction remains reserved for future work.

Currently, we assume that multiple applications can access *the same* signal class instance (i.e., the signal class instance with the same id), but among them, only one application has a right to *update* the persistent signals. Supporting a more generalized concurrent RP mechanism [15] also remains reserved for future work.

Persistent signals are also relevant for fault-tolerant RP [31, 32], which provides an implementation for the snapshotting mechanism of signals. Persistent signals also provide a basis for realizing such a fault-tolerant mechanism in that the execution can be reproductive because every update is recorded in the database. However, persistent signals focus more on applications that query over time-series data, such as IoT applications.

**Multi-tier RP languages**   Distribution of persistent signals implies the distribution of time-series data. Each persistent signal (or signal class instance) can be considered a micro-service that provides time-series data, and our proposal supports developing a new service or system by declaratively composing them. Actually, the knowledge already exists that signals are suitable for such a declarative composition in the context of a front-end application development [34]. This paper proposed a similar composition not in a front-end application but in one spreading over the network.

In this sense, our proposal is related to multi-tier languages that aid in implementing both a server and client using a single programming language. Several multi-tier languages such as Hop.js [41] and Links [13] do not support RP features. To our knowledge, integration of RP features with a multi-tier language first appeared in Scala Multi-Tier FRP [38]. This language is limited to the Web domain, and the FRP feature is supported by both the server-side and client-side APIs. SCALALOCI [46] is a multi-tier language that allows the specification of an architectural style of a distributed system, which consists of peers (different types of components) that are connected in different styles (e.g., a single tie or multiple ties). It can specify *placed data*, i.e., a value is placed in some peer. Communications between peers are represented using an RP style.

On the other hand, our proposal aims to support not the multi-tier feature but persistent data distribution in the distributed RP setting. Thus, our proposal does not distinguish tiers but every dataflow is specified by signals representing time-series data.

**Distributed recovery protocols**   Rollback-recovery ensures access to stable storage that survives all tolerated failures, and it has been intensively studied in the context of distributed systems, often viewed as message passing systems [18]. One significant challenge is to avoid rollback propagation, also known as the domino effect [37], where a failure in one process may trigger another process to rollback, thereby potentially causing a chain reaction of rollbacks. Coordinated checkpointing addresses this issue by allowing processes to take independent checkpoints [10].

In addition to checkpointing, the recovery method applied in our proposal bears similarities to log-based recovery [2, 45] in that we treat the update histories of source persistent signals as reliable logs. However, in the context of distributed persistent signals, rollbacks are not necessary; instead, we focus on re-executing computations after the latest checkpoint. Our formalization, particularly theorem 3.2, ensures that this re-execution maintains the time-series updates consistent with all data sources.

## 5  Conclusion and Future Work

This paper reports a design of reactive programming language with the support of imperative operations, transparent distribution of multiple data sources, and retroactive computation. A key challenge of this proposal is to develop a calculus to formally study the time consistency in this setting. This is because our language performs both implicit (reactive) computations and explicit ones. To unify them, we developed $\varsigma$-DPS, which is a hybrid of the object calculus (tailored to persistent signals) and a process calculus, and proved that the retroactive computation performed at each checkpoint actually ensures the consistency of each persistent signal with respect to all the data sources.

This study is a first step toward realizing declarative retroactive distributed dataflows and raises several interesting research questions, which should be answered in future work. First, as we designed the calculus in its simplest form, several extensions (e.g., relaxing the computation rules to enable propagation and other reductions (other than R-New and R-Setu) occur in parallel) can be considered. Next, it would be interesting to add transactions to the proposed declarative dataflows; i.e., making a couple of propagations an atomic operation. A similar idea already exists in the literature [44]; however, this feature was not studied in a declarative dataflow setting. Finally, we need to consider the effect of database schema evolution and refactoring of signal classes. A signal class and its corresponding database schema should co-evolve; i.e., a change in the database schema will require a change in the signal class declaration, and vice versa. Furthermore, old and new schema versions might be required to coexist;

this requirement is realistic, particularly given the open-ended settings discussed in this paper. However, supporting such an evolution while ensuring data consistency is a non-trivial issue. Currently, technologies that realize such database schema and persistent object evolution are being intensively studied [8, 20, 35]. We consider that the application of such technologies to our proposal will address this issue.

**Acknowledgements**   This study was supported by KAKENHI 21H03418 and 24K02922.

## A   Microbenchmarking

We confirm that the proposed language mechanism is realistic in several conceivable application scenarios by performing simple experiments. In particular, we set the following two research questions:

**RQ1.**  What overhead does the recovery process impose on the execution of the application?

**RQ2.**  How do the persistent signals consume storage space during the execution of the application?

For each research questions, we use the following application scenarios:

**Water level monitoring:**  Long spell of rain may cause river flooding. To predict the future water levels, this scenario continuously monitors the water levels and amounts of rain at some specified monitoring locations. The estimated future water level is calculated by a weighted sum of past values of water levels and rain amounts monitored at the surrounding monitoring locations. This monitoring is performed every 2 minutes. This is because, even though the Water Information System provided by the Ministry of Land, Infrastructure, Transport and Tourism, Japan[8] updates the water level information every 10 minutes, we consider more frequent updates might be appropriate in this scenario.

**Treadmill:**  The user of a treadmill system wants to observe statistics measured from the treadmill's belt and the runner's wearable device, such as estimated consumed calories, the current and average heart rate, and the exercise intensity, which is calculated by arithmetic operations using the average heartbeat rate and runner's age. As for the refresh rate of simple displays provided by many treadmill systems, we assume the update timing of each device to be one second.

**Traffic monitoring:**  This scenario provides network monitoring in the manner we mentioned above. While the traffic monitoring updates the traffic information every 5 minutes, the alive monitoring is triggered every minute. We consider these settings to be usual in network monitoring applications.

**Experimental settings**   To virtually run these scenarios, we prepared experimental settings, which are summarized as follows:

---

[8] http://www1.river.go.jp/ (visited on 2024-09-30).

■ **Table 1** Benchmark settings. We set the checkpoint intervals so that at least 5 update propagations are triggered between the intervals.

| Scenario | Checkpoint | @timing | Approx. trial time |
|---|---|---|---|
| **Waterlevel** | every 10 min | 2 min | 105 min |
| **Treadmill** | every 5 s | 1 s | 10 min |
| **Traffic** | every 5 min | 5 s (traffic), 1 min (alive) | 50 min |

■ **Table 2** Overhead of recovery process.

| | Total (ms) | # of invocations | Average time (ms) |
|---|---|---|---|
| Waterlevel | 39.0 | 11 | 3.55 |
| Treadmill | 466 | 240 | 1.94 |
| Traffic | 91.2 | 11 | 8.29 |

**Implementation:** The version of the source code for the proposed language's compiler, as well as its runtime library, used in this experiment is uploaded to http://hdl. handle.net/10559/0002013377. Notably, our implementation uses R2DBC (version 1.1.0, for PostgreSQL) to implement the reactive update propagation, as noted in Section 2.3.

**Sample scenarios:** We prepared sample mocks, rather than concrete applications, to virtually run the scenarios described above. The checkpoint and @timing settings are summarized in Table 1. These mocks are also uploaded to http://hdl.handle. net/10559/0002013377.

**Profiling method:** All scenarios are executed during the specified trial times, and all these executions are monitored using Java VisualVM 1.8.0_202 to obtain their profiling results.

**Storage consumption:** To evaluate the storage space consumption, we measured the actual values of the amount of storage space used by the database, because database tables are divided into several chunks (called hypertables) in TimescaleDB, and we could not trace the correspondence between tables and chunks appearing in the database system catalog (we performed a cleanup of the storage before each test run).

**Other settings:** All measurements were performed using PostgreSQL 14 with the TimescaleDB extension, which is running on six-cores Intel Xeon E-2276G 3.80GHz with 16GB main memory and 512GB SSD under Linux kernel version 4.18.0. The benchmark programs ran on six-cores AMD Ryzen 5 1600 3.20GHz with 8GB main memory under Linux kernel version 5.14.0.

**Evaluation regarding RQ1** Profiling results are summarized in Table 2, which shows the total execution times of the main method implementing the recovery process

■ **Table 3** Storage spaces consumed during the execution of applications. The waterlevel application did not require any extra pages during the test run.

|  | Waterlevel (105 min) | Treadmill (10 min) | Traffic (50 min) |
| --- | --- | --- | --- |
| Storage consumption (KB) | 0 (< 8) | 472 | 276 |

invoked for each checkpoint. This table indicates that the recovery process takes less than ten milliseconds for each checkpoint time. We note that this amount is less than 1% of the total CPU time. Because each checkpoint process may block the update of the underlying database, the overhead of the recovery process should be small. We consider that the measured overhead is small enough to continue normal executions without noticing the overhead, at least in the above scenarios where updates are performed every second or minute.

**Evaluation regarding RQ2**   The measured storage consumption is summarized in Table 3, which indicates that the storage space consumed during test runs is relatively small. This is partly because each update recorded in the database is minimal in size (e.g., no text data are used), and the update frequency is not high in these applications.

To interpret this result, we further need to estimate how these applications will consume storage space over long periods, as persistent signals preserve *all* past update records. Even though persistent signals can *theoretically* go back to the beginning of the application's lifecycle, operational limits may be imposed by archiving older data to save storage space. The appropriate limits vary according to the applications. For the treadmill application, we may want to preserve the record of a "30 minute excersize" (estimated at approximately 1.5MB) everyday for several years. For the traffic monitoring, it is also desirable to preserve the record for several years, and according to the test run, persistent signals require approximately 3GB of space per year. For the waterlevel monitoring, it may be necessary to provide analytics covering longer time spans, possibly several decades, and the test run shows that persistent signals will require approximately 35MB of storage space per year. These estimations indicate that storage consumption by persistent signals is not a problem for these applications.

## B   Examples

This section illustrates examples of process and explicit reductions introduced in Section 3.

### B.1  Example of Process Reductions

We consider the signal class network comprising the instance new $E(l_5$, new $C(l_3$, new $A(l_1)$, new $B(l_2))$, new $D(l_4))$, where signal class definitions are provided as follows (since the external sources only trigger the propagations (databases are not updated

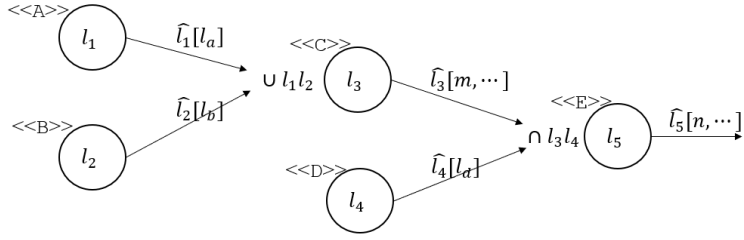

■ **Figure 8** Example of communication (i.e., update propagation) between processes.

externally) in ς-DPS, the most upstream persistent signals always emit the same value in the following example):

```
1  signal class A { persistent signal a=l_a; }
2  signal class B { persistent signal b=l_b; }
3  @mode("union") signal class C { signal c = m; a; b; ...}
4  signal class D { persistent signal d=l_d; }
5  @mode("intersection") signal class E { signal e = n; c; d; ...}
```

Using the syntax in Figure 4, this signal class network can be represented as the following process environment, which is also depicted in Figure 8:

$$\mathsf{l_1} \mapsto \hat{\mathsf{l_1}}[\mathsf{l}_a], \mathsf{l_2} \mapsto \hat{\mathsf{l_2}}[\mathsf{l}_b], \mathsf{l_3} \mapsto \cup \mathsf{l_1 l_2}.\hat{\mathsf{l_3}}[\mathsf{m}, \cdots], \mathsf{l_4} \mapsto \hat{\mathsf{l_4}}[\mathsf{l}_d], \mathsf{l_5} \mapsto \cap \mathsf{l_3 l_4}.\hat{\mathsf{l_5}}[\mathsf{n}, \cdots]$$

Assuming the update propagation is initiated by the external source referred by $\mathsf{l_2}$, the propagation is described as the following process reductions:

$$\mu \mid \mathsf{l_1} \mapsto \hat{\mathsf{l_1}}[\mathsf{l}_a], \mathsf{l_2} \mapsto \hat{\mathsf{l_2}}[\mathsf{l}_b], \mathsf{l_3} \mapsto \cup \mathsf{l_1 l_2}.\hat{\mathsf{l_3}}[\mathsf{m}, \cdots], \mathsf{l_4} \mapsto \hat{\mathsf{l_4}}[\mathsf{l}_d], \mathsf{l_5} \mapsto \cap \mathsf{l_3 l_4}.\hat{\mathsf{l_5}}[\mathsf{n}, \cdots]$$

$\rightsquigarrow \mu \mid_{\mathsf{l_2}} \mathsf{l_1} \mapsto \hat{\mathsf{l_1}}[\mathsf{l}_a], \mathsf{l_2} \mapsto \hat{\mathsf{l_2}}, \mathsf{l_3} \mapsto \cup \mathsf{l_1 l_2}.\hat{\mathsf{l_3}}[\mathsf{m}, \cdots], \mathsf{l_4} \mapsto \hat{\mathsf{l_4}}[\mathsf{l}_d], \mathsf{l_5} \mapsto \cap \mathsf{l_3 l_4}.\hat{\mathsf{l_5}}[\mathsf{n}, \cdots]$
    (initiating the propagation by unguarding $\mathsf{l_2}$)

$\rightsquigarrow_{\mathsf{PR\text{-}OR}} \mu' \mid_{\mathsf{l_3,l_2}} \mathsf{l_1} \mapsto \hat{\mathsf{l_1}}[\mathsf{l}_a], \mathsf{l_2} \mapsto \hat{\mathsf{l_2}}, \mathsf{l_3} \mapsto \hat{\mathsf{l_3}}, \mathsf{l_4} \mapsto \hat{\mathsf{l_4}}[\mathsf{l}_d], \mathsf{l_5} \mapsto \cap \mathsf{l_3 l_4}.\hat{\mathsf{l_5}}[\mathsf{n}, \cdots]$
    (evaluating $\mathsf{m}$ to update the field $\mathsf{c}$ of $\mathsf{l_3}$ and updating $\mu$)

$\rightsquigarrow \mu' \mid_{\mathsf{l_4,l_3,l_2}} \mathsf{l_1} \mapsto \hat{\mathsf{l_1}}[\mathsf{l}_a], \mathsf{l_2} \mapsto \hat{\mathsf{l_2}}, \mathsf{l_3} \mapsto \hat{\mathsf{l_3}}, \mathsf{l_4} \mapsto \hat{\mathsf{l_4}}, \mathsf{l_5} \mapsto \cap \mathsf{l_3 l_4}.\hat{\mathsf{l_5}}[\mathsf{n}, \cdots]$
    (unguarding $\mathsf{l_4}$)

$\rightsquigarrow_{\mathsf{PR\text{-}AND}} \mu'' \mid_{\mathsf{l_5,l_4,l_3,l_2}} \mathsf{l_1} \mapsto \hat{\mathsf{l_1}}[\mathsf{l}_a], \mathsf{l_2} \mapsto \hat{\mathsf{l_2}}, \mathsf{l_3} \mapsto \hat{\mathsf{l_3}}, \mathsf{l_4} \mapsto \hat{\mathsf{l_4}}, \mathsf{l_5} \mapsto \hat{\mathsf{l_5}}$
    (evaluating $\mathsf{n}$ to update the field $\mathsf{e}$ of $\mathsf{l_5}$ and updating $\mu'$)

$\rightsquigarrow \mu'' \mid_{\mathsf{l_1,l_5,l_4,l_3,l_2}} \mathsf{l_1} \mapsto \hat{\mathsf{l_1}}, \mathsf{l_2} \mapsto \hat{\mathsf{l_2}}, \mathsf{l_3} \mapsto \hat{\mathsf{l_3}}, \mathsf{l_4} \mapsto \hat{\mathsf{l_4}}, \mathsf{l_5} \mapsto \hat{\mathsf{l_5}}$
    (unguarding $\mathsf{l_1}$, resulting in a normal form)

### B.2 Example of the Whole Reductions

We exemplify how process reductions and explicit reductions are merged in ς-DPS. Assuming the process environment depicted in Figure 8 (denoted as $\nu$ in the following example), computations of the expression $\mathsf{l_5}.\mathsf{setu}(\mathsf{l_3}, \mathsf{l_6}[\mathsf{d} = \mathsf{l}_{d'}])$ at time $\mathsf{t}$ proceed as follows:

$\mu; \nu; \phi; t \mid l_5.\text{setu}(l_3, l_6[d = l_{d'}])$
$\quad \longrightarrow_{\text{R-Obj}} \mu'; \nu \oplus l_6 \mapsto l_6[l_{d'}]; \phi'; t \mid l_5.\text{setu}(l_3, l_6)$
$\qquad (\text{where } \phi' = \phi \cup \{(t, \nu \oplus l_6 \mapsto l_6[l_{d'}]\})$
$\quad \longrightarrow_{\text{PROPAGATION}} \mu'; \nu; \phi; t \mid l_5.\text{setu}(l_3, l_6[d = l_{d'}])$
$\qquad (\text{if updates are found in sources, propagating them})$
$\quad \rightarrow_{\text{R-Step}} \mu'; \nu \oplus l_6 \mapsto l_6[l_{d'}]; \phi'; t + 1 \mid l_5.\text{setu}(l_3, l_6)$
$\qquad (\text{incrementing time})$
$\quad \longrightarrow_{\text{R-Setu}} \mu''; (\nu \oplus l_6 \mapsto l_6[l_{d'}]) \oplus l_5 \mapsto \cap l_3 l_6.\hat{l_5}[m]; \phi''; t + 1 \mid l_5$
$\qquad (\text{where } \phi'' = \phi' \cup \{(t+1, (\nu \oplus l_6 \mapsto l_6[l_{d'}]) \oplus l_5 \mapsto \cap l_3 l_6.\hat{l_5}[m])\})$
$\quad \longrightarrow_{\text{PROPAGATION}} \mu''; \nu \oplus l_6 \mapsto l_6[l_{d'}]; \phi'; t + 1 \mid l_5.\text{setu}(l_3, l_6)$
$\qquad (\text{if updates are found in sources, propagating them})$
$\quad \rightarrow_{\text{R-Step}} \mu''; (\nu \oplus l_6 \mapsto l_6[l_{d'}]) \oplus l_5 \mapsto \cap l_3 l_6.\hat{l_5}[m]; \phi''; t + 2 \mid l_5$
$\qquad (\text{incrementing time})$

As mentioned before, each R-STEP computes zero or more update propagations in actual time occurred between explicit reductions, and (logical) time in our calculus only proceeds by R-STEP. Without loss of generality, this abstraction of time simplifies the computation rules in $\varsigma$-DPS and discussions regarding time consistency below.

## C  Proofs

This section shows the proofs of theorems introduced in Section 3.

### C.1  Notation

Let

- $\iota(p)$ be the input channels of the process $p$
- $\mathscr{R}(\mu(l))$ be the data source $\mathscr{R}_l$ of $\mu(l) = (\mathscr{R}_l, tm, \odot)$
- $mode(\mu(l))$ be the mode $\odot$ of $\mu(l) = (\mathscr{R}_l, tm, \odot)$
- $tm(\mu(l))$ be the update time $tm$ of $\mu(l) = (\mathscr{R}_l, tm, \odot)$
- $\mu \overset{\mathscr{R}^{\text{cond}}}{=} \mu'$ if $\forall l \in dom(\mu). \forall t.\text{cond}(t) \implies \exists \bar{l}.((t, \bar{l}) \in \mathscr{R}(\mu(l)) \iff (t, \bar{l}) \in \mathscr{R}(\mu'(l)))$
- $\mu_{\bar{l}}$ restricts the domain of $\mu$ to $\bar{l}$, i.e., $\mu_{\bar{l}} = \{(l, \mu(l)) \mid l \in \bar{l}\}$
- $\mu \overset{tm}{=} \mu'$ if $\forall l \in dom(\mu).tm(\mu(l)) = tm(\mu(l'))$
- $\mu \overset{mode}{=} \mu'$ if $\forall l \in dom(\mu).mode(\mu(l)) = mode(\mu(l'))$
- $dom_{\text{cond}}(\phi) = \{t \in dom(\phi) \mid \text{cond}(t)\}$
- $\phi \overset{\text{cond}}{=} \phi'$ if $dom_{\text{cond}}(\phi) = dom_{\text{cond}}(\phi')$ and $\forall t \in dom_{\text{cond}}(\phi).\phi(t) = \phi'(t)$

## C.2 Consistent Relation

$\mu$ is consistent under $v$ at $\mathsf{t}$ and $\mathsf{l}$ if

1. If $mode(\mu(\mathsf{l})) = \cap$:

$$\exists \vec{\mathsf{l}}.(t,\vec{\mathsf{l}}) \in \mathscr{R}(\mu(\mathsf{l})) \iff \forall \mathsf{l}_i \in \iota(v(\mathsf{l})).\exists \vec{\mathsf{l}'}.(t,\vec{\mathsf{l}'}) \in \mathscr{R}(\mu(\mathsf{l}_i))$$

2. If $mode(\mu(\mathsf{l})) = \cup$:

$$\exists \vec{\mathsf{l}}.(t,\vec{\mathsf{l}}) \in \mathscr{R}(\mu(\mathsf{l})) \iff \exists \mathsf{l}_i \in \iota(v(\mathsf{l})).\exists \vec{\mathsf{l}'}.(t,\vec{\mathsf{l}'}) \in \mathscr{R}(\mu(\mathsf{l}_i))$$

## C.3 Multistep Pure Reduction

$$\mu; v; \mathsf{t} \vdash \mathsf{e} \longrightarrow^0 \mathsf{e} \qquad\qquad \text{(R-M-Zero)}$$

$$\frac{\mu; v; \mathsf{t} \vdash \mathsf{e} \longrightarrow^n \mathsf{e}' \qquad \mu; v; t \vdash \mathsf{e}' \longrightarrow \mathsf{e}''}{\mu; v; \mathsf{t} \vdash \mathsf{e} \longrightarrow^{n+1} \mathsf{e}''} \qquad\qquad \text{(R-M-One)}$$

We write $\longrightarrow^*$ if the number of steps is not important.

## C.4 Multistep Process Reduction

$$v; \mathsf{t} \vdash \mu \mid_{\vec{\mathsf{l}}} v' \rightsquigarrow^0 \mu \mid_{\vec{\mathsf{l}}} v' \qquad\qquad \text{(Pr-M-Zero)}$$

$$\frac{v; \mathsf{t} \vdash \mu \mid_{\vec{\mathsf{l}}} v' \rightsquigarrow^n \mu' \mid_{\vec{\mathsf{l}'}} v'' \qquad v; \mathsf{t} \vdash \mu' \mid_{\vec{\mathsf{l}'}} v'' \rightsquigarrow \mu'' \mid_{\mathsf{l},\vec{\mathsf{l}'}} v'''}{v; \mathsf{t} \vdash \mu \mid_{\vec{\mathsf{l}}} v' \rightsquigarrow^{n+1} \mu'' \mid_{\mathsf{l},\vec{\mathsf{l}'}} v'''} \quad \text{(Pr-M-One)}$$

## C.5 Lemmas for Theorem 3.1

**Lemma C.1.** *Suppose that $v''; \mathsf{t} \vdash \mu \mid_{\vec{\mathsf{l}}} v \rightsquigarrow \mu' \mid_{\vec{\mathsf{l}'}} v'$. Then,*

1. $\mu \overset{tm}{=} \mu'$
2. $\mu \overset{mode}{=} \mu'$
3. $\mu \overset{\mathscr{R}^{<t}}{=} \mu'$

*Proof.* By case analysis on the derivation of $v''; \mathsf{t} \vdash \mu \mid_{\vec{\mathsf{l}}} v \rightsquigarrow \mu' \mid_{\vec{\mathsf{l}'}} v'$. □

**Lemma C.2.** *Suppose that $v''; \mathsf{t} \vdash \mu \mid_{\vec{\mathsf{l}}} v \rightsquigarrow^n \mu' \mid_{\vec{\mathsf{l}'}} v'$. Then,*

1. $\mu \overset{tm}{=} \mu'$
2. $\mu \overset{mode}{=} \mu'$
3. $\mu \overset{\mathscr{R}^{<t}}{=} \mu'$

*Proof.* By Lemma C.1 and induction on the derivation of $v''; \mathsf{t} \vdash \mu \mid_{\vec{\mathsf{l}}} v \rightsquigarrow^n \mu' \mid_{\vec{\mathsf{l}'}} v'$. □

**Lemma C.3.** *Suppose $\nu; t \vdash \mu \longrightarrow_{\bar{l}} \mu'$. Then,*

1. $\mu \stackrel{tm}{=} \mu'$
2. $\mu \stackrel{mode}{=} \mu'$
3. $\mu \stackrel{\mathcal{R}^{<t}}{=} \mu'$

*Proof.* By Lemma C.2. □

**Lemma C.4.** *Assume that for given $\bar{l}, \mu, \nu, \mu'$, and $\nu'$,*

- *$\nu$ is wellformed under $\mu$ i.e., $\forall l \in dom(\mu).\mu, \nu \vdash l$ OK*

*and $dom(\nu) \subseteq dom(\mu)$*

- *$\forall l \in \bar{l}.\mu$ is consistent under $\nu$ at $t$ and $l$*
- *$dom(\nu) = dom(\nu')$*
- *$\forall l \in dom(\mu) \setminus \bar{l}.\nu(l) = \nu'(l)$*
- *$\forall l \in \bar{l}.\exists \bar{l'}.(t, \bar{l'}) \in \mathcal{R}(\mu(l))$*
- *$\forall l \in \bar{l}.\nu'(l) = \hat{l}$ and $t \equiv 0 \mod tm(\mu(l))$*
- *$\nu; t \vdash \mu' \mid_{\bar{l}} \nu' \rightsquigarrow \mu' \mid_{l,\bar{l}} \nu''$*

*Then we have*

1. *$dom(\nu'') \subseteq dom(\mu')$*
2. *$\forall l \in dom(\nu) \setminus \{l\}.\nu'(l') = \nu''(l')$*
3. *$\forall l' \in (l, \bar{l}).\exists \bar{l''}.(t, \bar{l''}) \in \mathcal{R}(\mu'(l'))$*
4. *$\forall l' \in (l, \bar{l}).\nu''(l') = \hat{l}$ and $t \equiv 0 \mod tm(\mu'(l'))$*
5. *$\forall l' \in (l, \bar{l}).\mu'$ is consistent under $\nu$ at $t$ and $l'$*
6. *$\nu$ is well-formed under $\mu'$*

*Proof.* By case analysis on derivation of $\nu; t \vdash \mu \mid_{\bar{l}} \nu' \rightsquigarrow \mu' \mid_{l,\bar{l}} \nu''$.

- PR-AND.
  - The properties 1 and 2 hold because the process reduction just updates the values for the key $l$ in $\mu$ and $\nu'$.
  - The property 3. holds because
    * $\mu(l') = \mu'(l')$ for all $l' \in \bar{l}$
    * assumption: $\forall l \in \bar{l}.\exists \bar{l'}.(t, \bar{l'}) \in \mathcal{R}(\mu(l))$
    * $\mathcal{R}(\mu'(l)) = \mathcal{R}(\mu(l)) \oplus l \mapsto (\mathcal{R}(\mu(l)) \oplus \{(t, \dots)\})$.
  - The property 4 holds because
    * assumption: $\forall l \in \bar{l}.\nu'(l) = \hat{l}$ and $t \equiv 0 \mod tm(\mu(l))$
    * the process indexed by $l$ in $\nu'$ is just reduced to $l$, i.e., $\nu''(l) = \hat{l}$ if $t \equiv 0 \mod tm(\mu(l))$
  - The property 5 holds because
    * $\mu'$ is consistent under $\nu$ at $t$ for all $l' \in \bar{l}$ because $\mu(l') = \mu'(l')$
    * $\mu'$ is consistent under $\nu$ at $t$ and $l$ because $\iota(\nu(l)) \subseteq \bar{l}$

- The property 6 holds because
  * $dom(\mu) = dom(\mu')$
  * Lemma C.1
  * $v$ is well-formed under $\mu$
- Pr-Or. Similar to Pr-And.

$\square$

**Lemma C.5.** *Assume that*

- *$v$ is wellformed under $\mu$*
- *$\forall l \in \bar{l}.\mu$ is consistent under $v$ at $t$ and $l$*
- *$dom(v) = dom(v')$*
- *$\forall l \in dom(\mu) \setminus \bar{l}.v(l) = v'(l)$*
- *$\forall l \in \bar{l}.\exists \overline{l'}.(t,\overline{l'}) \in \mathscr{R}(\mu(l))$*
- *$\forall l \in \bar{l}.v'(l) = \hat{l}$ and $t \equiv 0 \mod tm(\mu(l))$*
- *$v;t \vdash \mu \mid_{\bar{l}} v' \rightsquigarrow^* \mu' \mid_{\bar{l'}} v''$*

*Then we have*

1. $dom(v'') \subseteq dom(\mu')$
2. $\bar{l} \subseteq \overline{l'}$
3. $\forall l \in dom(v) \setminus \overline{l'}.v'(l') = v''(l')$
4. $\forall l' \in \overline{l'}.\exists \overline{l''}.(t,\overline{l''}) \in \mathscr{R}(\mu'(l'))$
5. $\forall l' \in \overline{l'}.v''(l') = \hat{l'}$ and $t \equiv 0 \mod tm(\mu(l'))$
6. $\forall l' \in \overline{l'}.\mu'$ is consistent under $v$ at $t$ and $l$
7. $v$ is well-formed under $\mu'$

*Proof.* By Lemma C.4 and induction on derivation of $v;t \vdash \mu \mid_{\bar{l}} v' \rightsquigarrow^* \mu' \mid_{l,\bar{l}} v''$. $\square$

**Lemma C.6.** *Assume that*

- *$v$ is wellformed under $\mu$*
- *$\forall l \in dom(\mu).(\exists \overline{l'}.(t,\overline{l'}) \in \mathscr{R}(\mu(l))) \implies t \equiv 0 \mod tm(\mu(l))$*
- *$v;t \vdash \mu \longrightarrow_{\bar{l}} \mu'$*

*Then $\mu'$ is consistent under $v$ at $t$.*

*Proof.* From the definition of Propagation and Lemma C.5, we have

- $v;t \vdash \mu \mid_{\emptyset} v \rightsquigarrow^* \mu' \mid_{\bar{l}} v'$
- $dom(v) = dom(v')$
- $\mu(l) = \mu'(l)$ and $v(l) = v'(l)$ for all $l \in dom(\mu) \setminus \bar{l}$
- $\forall l' \in \bar{l}.\exists \overline{l'}.(t,\overline{l'}) \in \mathscr{R}(\mu'(l))$
- $\forall l \in \bar{l}.v'(l) = \hat{l}$ and $t \equiv 0 \mod tm(\mu(l))$
- $\forall l \in \bar{l}.\mu'$ is consistent under $v$ at $t$ and $l$
- $\bar{l} = \{l \in dom(\mu) \mid t \equiv 0 \mod tm(\mu(l))\}$

$\mu(l)$ may contain a record at time $t$ for any $l \in dom(\mu)$, but they are all overridden in $\mu'$. Therefore $\mu'$ is consistent under $v$ at $t$. $\square$

**Lemma C.7.** *Suppose that $\mu; t \vdash \nu; \phi \mid e \longrightarrow \nu'; \phi' \mid e'$. Then $\phi \stackrel{\leq t}{=} \phi'$.*

*Proof.* By induction on derivation of $\mu; t \vdash \nu; \phi \mid e \longrightarrow \nu'; \phi' \mid e'$. □

**Lemma C.8.** *Suppose that*
- $\phi(t) = \nu$
- $\mu; t \vdash \nu; \phi \mid e \longrightarrow \nu'; \phi' \mid e'$

*Then $\phi'(t) = \nu'$.*

*Proof.* By induction on derivation of $\mu; t \vdash \nu; \phi \mid e \longrightarrow \nu'; \phi' \mid e'$. □

**Lemma C.9.** *Assume that*
- $\nu$ *is well-formed under* $\mu$
- $\mu; t \vdash \nu; \phi \mid e \longrightarrow \nu'; \phi' \mid e'$

*Then $\nu'$ is well-formed under $\mu$.*

*Proof.* By induction on derivation of $\mu; t \vdash \nu; \phi \mid e \longrightarrow \nu'; \phi' \mid e'$. □

## C.6 Proof of Theorem 3.1

*Proof.* R-STEP gives us
1. $\mu; t \vdash \nu; \phi \mid e \longrightarrow \nu'; \phi' \mid e'$
2. $\nu'; t \vdash \mu \longrightarrow \mu'$

By Lemmas C.7, C.8 and C.9, we have
- $\nu'$ is well-formed under $\mu$
- $\phi \stackrel{\leq t}{=} \phi'$
- $\phi'(t) = \nu'$

By Lemma C.6, $\mu'$ is consistent under $\nu'$ at t. Because $\phi'(t) = \nu'$, $\mu'$ is consistent under $\phi'(t)$ at t. □

## C.7 Lemmas for Theorem 3.2

**Lemma C.10.** *Suppose that*
- $\nu(l) = \odot \vec{l'}.\hat{l}[\vec{e}, \cdots]$
- $\mu^1 \stackrel{tm}{=} \mu^2$
- $\mu^1 \stackrel{mode}{=} \mu^2$
- $\mu^1 \stackrel{\mathscr{R}^{<t}}{=} \mu^2$
- $\mu^1_l \stackrel{\mathscr{R}^{=t}}{=} \mu^2_l$
- $\mu^1; \nu; t \vdash_{p(\hat{l})} e_i \longrightarrow e'_i$

*Then $\mu^2; \nu; t \vdash_{p(\hat{l})} e_i \longrightarrow e'_i$.*

*Proof.* By induction on derivation of $\mu^1; \nu; t \vdash_{p(\hat{l})} e \longrightarrow e'$. □

**Lemma C.11.** *Suppose that*

- $\nu(l) = \odot \overline{l'}.\hat{l}[\overline{e}, \cdots]$
- $\mu^1 \overset{tm}{=} \mu^2$
- $\mu^1 \overset{mode}{=} \mu^2$
- $\mu^1 \overset{\mathscr{R}^{<t}}{=} \mu^2$
- $\mu^1_{\hat{l}} \overset{\mathscr{R}^{=t}}{=} \mu^2_{\hat{l}}$
- $\mu^1; \nu; t \vdash_{p(\hat{l})} e \longrightarrow^n e'$

    *Then* $\mu^2; \nu; t \vdash_{p(\hat{l})} e \longrightarrow^n e'$

*Proof.* By Lemma C.10 and induction on derivation of $\mu^1; \nu; t \vdash_{p(\hat{l})} e \longrightarrow^n e'$. $\qquad\square$

**Lemma C.12.** *Suppose that*

- $\mu^1 \overset{tm}{=} \mu^3$
- $\mu^1 \overset{mode}{=} \mu^3$
- $\mu^1 \overset{\mathscr{R}^{<t}}{=} \mu^3$
- $\mu^1_{\hat{l}} \overset{\mathscr{R}^{=t}}{=} \mu^3_{\hat{l}}$
- $\nu; t \vdash \mu^1 \mid_{\hat{l}} \nu^1 \rightsquigarrow \mu^2 \mid_{l,\hat{l}} \nu^2$

    *Then there exists* $\mu^4$ *such that*

- $\nu; t \vdash \mu^3 \mid_{\hat{l}} \nu^1 \rightsquigarrow \mu^4 \mid_{l,\hat{l}} \nu^2$
- $\mu^2 \overset{tm}{=} \mu^4$ *and* $\mu^2 \overset{mode}{=} \mu^4$
- $\mu^2 \overset{\mathscr{R}^{<t}}{=} \mu^4$
- $\mu^2_{l,\hat{l}} \overset{\mathscr{R}^{=t}}{=} \mu^4_{l,\hat{l}}$

*Proof.* By case analysis on the derivation of $\nu''; t \vdash \mu \mid_{\hat{l}} \nu \rightsquigarrow \mu' \mid_{l,\hat{l}} \nu'$. $\qquad\square$

**Lemma C.13.** *Suppose that*

- $\mu^1 \overset{tm}{=} \mu^3$
- $\mu^1 \overset{mode}{=} \mu^3$
- $\mu^1 \overset{\mathscr{R}^{<t}}{=} \mu^3$
- $\mu^1_{\hat{l}} \overset{\mathscr{R}^{=t}}{=} \mu^3_{\hat{l}}$
- $\nu; t \vdash \mu^1 \mid_{\hat{l}} \nu^1 \rightsquigarrow^n \mu^2 \mid_{\overline{l'}} \nu^2$

    *Then there exists* $\mu^4$ *such that*

- $\nu; t \vdash \mu^3 \mid_{\hat{l}} \nu^1 \rightsquigarrow^n \mu^4 \mid_{\overline{l'}} \nu^2$
- $\mu^2 \overset{tm}{=} \mu^4$ *and* $\mu^2 \overset{mode}{=} \mu^4$
- $\mu^2 \overset{\mathscr{R}^{<t}}{=} \mu^4$
- $\mu^2_{\overline{l'}} \overset{\mathscr{R}^{=t}}{=} \mu^4_{\overline{l'}}$

*Proof.* By Lemma C.12 and induction on the derivation of $\nu; t \vdash \mu^1 \mid_{\hat{l}} \nu^1 \rightsquigarrow^n \mu^2 \mid_{\hat{l}} \nu^2$. $\qquad\square$

**Lemma C.14.** *Suppose that*

- $\mu^1 \stackrel{tm}{=} \mu^3$
- $\mu^1 \stackrel{mode}{=} \mu^3$
- $\mu^1 \stackrel{\mathscr{R}^{<t}}{=} \mu^3$
- $\nu; t \vdash \mu^1 \longrightarrow_{\bar{i}} \mu^2$

  *Then there exists $\mu^4$ such that*
- $\nu; t \vdash \mu^3 \longrightarrow_{\bar{i}} \mu^4$
- $\mu^2 \stackrel{tm}{=} \mu^4$
- $\mu^2 \stackrel{mode}{=} \mu^4$
- $\mu^2 \stackrel{\mathscr{R}^{<t}}{=} \mu^4$
- $\mu^2_{\bar{i}} \stackrel{\mathscr{R}^{=t}}{=} \mu^4_{\bar{i}}$

*Proof.* By Lemma C.13. □

**Lemma C.15.** *Suppose that $\nu; t \vdash \mu \longrightarrow_{\bar{i}} \mu'$. Then*

- $\mu \stackrel{\mathscr{R}^{<t}}{=} \mu'$
- $\mu \stackrel{tm}{=} \mu'$
- $\mu \stackrel{mode}{=} \mu'$

*Proof.* By Lemma C.2. □

**Lemma C.16.** *Suppose that*

- $\mu^1 \stackrel{\mathscr{R}^{<t}}{=} \mu^2$
- $\mu^1 \stackrel{tm}{=} \mu^2$
- $\mu^1 \stackrel{mode}{=} \mu^2$
- $\phi^3 \stackrel{\leq t}{=} \phi^2$
- $\mu^1; t \vdash \nu; \phi^1 \mid e \longrightarrow \nu'; \phi^3 \mid e'$

  *Then there exists $\nu'', \phi^4, e''$ such that*
- $\mu^2; t \vdash \nu; \phi^2 \mid e \longrightarrow \nu''; \phi^4 \mid e''$
- $\nu' = \nu''$
- $\phi^3 \stackrel{\leq t}{=} \phi^4$
- $e' = e''$

*Proof.* By induction on the derivation of $\mu^1; t \vdash \nu; \phi^1 \mid e \longrightarrow \nu'; \phi^3 \mid e'$. □

Below, we prove a lemma more general than Theorem 3.2 Then we will prove Theorem 3.2 as a corollary of the lemma.

**Lemma C.17.** *Suppose that*

- $\forall t' \in dom(\phi).t' < t$
- $\phi(t) = \nu$
- $\mu; \nu; \phi; t \mid e \twoheadrightarrow_{\bar{i}} \mu^1; \nu^1; \phi^1; t^1 \mid e'$

- $\mu; \nu; \phi; t \mid e \twoheadrightarrow_{\overline{l'}} \mu^2; \nu^2; \phi^2; t^2 \mid e''$
- $dom(\mu^2) = dom(\mu^3)$
- $\mu^2 \stackrel{\mathscr{R}^{\leq t}}{=} \mu^3$
- $\mu^2 \stackrel{tm}{=} \mu^3$
- $\mu^2 \stackrel{mode}{=} \mu^3$
- $\phi^2 \stackrel{\leq t}{=} \phi^3$

  Then there exists $\mu^4, \nu^4, \phi^4, t^4, e'''$ such that

- $\mu^3; \phi^3(t-1); \phi^3; t \mid e \twoheadrightarrow_{\overline{l}} \mu^4; \nu^4; \phi^4; t^4 \mid e'''$
- $\mu^1 \stackrel{\mathscr{R}^{<t}}{=} \mu^4$
- $\mu^1_{\overline{l}} \stackrel{\mathscr{R}^{=t}}{=} \mu^4_{\overline{l}}$
- $\mu^1 \stackrel{tm}{=} \mu^4$
- $\mu^1 \stackrel{mode}{=} \mu^4$
- $\nu^1 = \nu^4$
- $\phi^1 \stackrel{\leq t}{=} \phi^4$
- $e' = e'''$
- $t^1 = t^4$

*Proof.* $t^1 = t^4$ is obvious. We can derive that

- $\phi(t) = \nu = \phi(t-1)$ from the first and second assumptions
- $\phi^1 = \phi^2$ from the definitions of R-Step and R-StepLost, noticing that the results of the reduction parts are the same in the both rules and $\phi^1$ and $\phi^2$ do not change in the propagation parts.
- $\phi \stackrel{\leq t}{=} \phi^1$ from Lemma C.7.
- $\mu \stackrel{\mathscr{R}^{<t}}{=} \mu^1 \stackrel{\mathscr{R}^{<t}}{=} \mu^2$, $\mu \stackrel{mode}{=} \mu^1 \stackrel{mode}{=} \mu^2$, and $\mu \stackrel{tm}{=} \mu^1 \stackrel{tm}{=} \mu^2$ from Lemmas C.3 and C.15.

  Therefore $\phi^3(t-1) = \phi^2(t-1) = \phi(t-1) = \nu$, which allows us to use Lemmas C.14 and C.16 to obtain the remaining properties. □

### C.8 Proof of Theorem 3.2

*Proof.* By Lemma C.17. $\overline{l'}$ is always a subset (but may not be a subsequence) of $\overline{l}$. This ensures that $\forall l \in dom(\mu^2) \setminus \overline{l}.\ \nexists \overline{l''}.(t, \overline{l''}) \in \mathscr{R}(\mu^2(l))$; therefore, for all $l$ in $dom(\mu^3) \setminus \overline{l}$, $\nexists \overline{l''}.(t, \overline{l''}) \in \mathscr{R}(\mu^3(l))$. Because $\forall l \in dom(\mu^1) \setminus \overline{l}.\ \nexists \overline{l''}.(t, \overline{l''}) \in \mathscr{R}(\mu^1(l))$, $\mu^1 \stackrel{\mathscr{R}^{<t}}{=} \mu^3$, and $\mu^1_{\overline{l}} \stackrel{\mathscr{R}^{=t}}{=} \mu^3_{\overline{l}}$, we obtain $\mu^1 \stackrel{\mathscr{R}^{\leq t}}{=} \mu^3$. □

## About the authors

**Tetsuo Kamina** is an Associate Professor at Oita University. His research interests include the design and implementation of programming languages, programming practices, and programming education. You can contact him at kamina@oita-u.ac.jp.
https://orcid.org/0000-0003-0288-1908

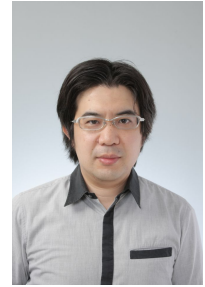

**Tomoyuki Aotani** is an Associate Professor at Sanyo Onoda City University. His research interests focus on the principles and mechanisms of programming languages, with a particular emphasis on modularity. You can contact him at aotani@rs.socu.ac.jp.
https://orcid.org/0000-0003-4538-0230

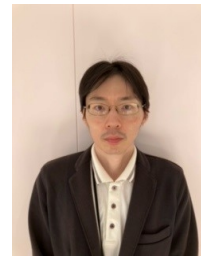

**Hidehiko Masuhara** is a Professor of Mathematical and Computing Science at Institute of Science Tokyo (formerly Tokyo Institute of Technology). His research interest is programming languages, especially on aspect- and context-oriented programming, partial evaluation, computational reflection, meta-level architectures, parallel/concurrent computing, and programming environments. Contact him at masuhara@acm.org.
https://orcid.org/0000-0002-8837-5303

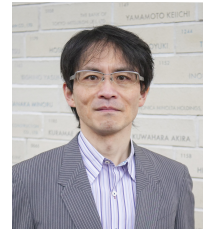